\documentclass[a4paper,11pt]{article}
\pdfoutput=1 

\usepackage{jcappub} 
\usepackage{natbib} 

\usepackage[T1]{fontenc} 

\usepackage{amsmath}
\usepackage{amsfonts}
\usepackage{amssymb}
\usepackage{bm}
\usepackage{physics}
\usepackage{graphicx}
\usepackage{subcaption}

\usepackage{multicol,multirow}
\usepackage{booktabs}
\usepackage{array}
\usepackage{appendix}
\usepackage{xcolor}
\usepackage{soul}
\usepackage{colortbl}
\usepackage{orcidlink}

\title{\boldmath Dark Energy Bubble as Dynamical Dark Energy:\\
Properties and CMB Constraints}

\author[a,b,1]{Batia Friedman-Shaw\orcidlink{0009-0008-1840-8481},}
\author[a,b,c]{Matthew C. Johnson\orcidlink{0000-0003-4199-0314},}
\author[a,b,c]{Katherine J. Mack\orcidlink{0000-0001-8927-1795}}

\affiliation[a]{Perimeter Institute for Theoretical Physics,\\
31 Caroline St N, Waterloo, Ontario, N2L 2Y5, Canada}
\affiliation[b]{Department of Physics and Astronomy, University of Waterloo,\\
200 University Ave W, Waterloo, Ontario, N2L 3G1, Canada}
\affiliation[c]{Department of Physics and Astronomy, York University,
Toronto, Ontario, M3J 1P3, Canada}

\emailAdd{bfriedmanshaw@perimeterinstitute.ca}
\emailAdd{mjohnson@perimeterinstitute.ca}
\emailAdd{kmack@perimeterinstitute.ca}

\note{Corresponding author.}

\abstract{Recent DESI results are in tension with the constant dark energy density predicted by the $\Lambda$CDM model. If dark energy is associated with the vacuum energy of a scalar field in a metastable state, it will undergo a first-order phase transition through the nucleation of bubbles containing a reduced dark energy density. In this paper, we explore the consequences of this model, where dark energy density varies in both space and time. We model a single bubble spacetime using the Israel junction conditions and derive each of the usual distance measures in this inhomogeneous cosmology. We find that the model predictions of Alcock--Paczynski distortion have features that align strikingly well with the DESI measurements if the dark energy phase transition occurred at roughly a redshift of 1.4 and if the bubble of lower dark energy density has roughly 10\% less dark energy density than the outer cosmology. Despite this feature, we find that the dark energy bubble is heavily constrained by the CMB which excludes the region of parameter space that reproduces the DESI BAO measurements. Still, the peculiar features in the distance measurements of the dark energy bubble cosmology serve as a useful toy model to motivate and inform future work in the currently poorly explored area of spatially varying dynamical dark energy.}

\begin{document}
\maketitle
\section{Introduction}

\label{sec:intro}

Modern cosmological observations, including spectroscopic galaxy~\cite{DESI_DR1, DESI_DR2} and Cosmic Microwave Background (CMB) \cite{act_dr6, Planck_2018_cosmo, Planck_2018_dipole, SPT} surveys, are now testing the standard $\Lambda$CDM model with unprecedented precision. In particular, Dark Energy Spectroscopic Instrument (DESI) Baryon Acoustic Oscillation (BAO) measurements have sharpened constraints on cosmic expansion and, when combined with CMB and supernova data, show a preference for extensions of $\Lambda$CDM in which the dark energy equation of state evolves with redshift \cite{DESI_DR1, DESI_DR2, Planck_2018_cosmo, DESY5, Union3, Pantheon+}. A common phenomenological description for this behaviour is the Chevallier--Polarski--Linder (CPL) parametrization, $w(a)=w_0+w_a(1-a)$ \cite{Chevallier_2001, Linder_2003}, which is usually interpreted as homogeneous evolution of the dark energy equation of state. However, because cosmological surveys are conducted in redshift space, an apparent redshift dependence could also be explained by spatial variation along the past light cone. \footnote{This possibility is similar to local-void and other radially inhomogeneous cosmologies, where spatial structure can affect the inferred distance-redshift relation~\cite{Marra_2013,Garcia_Bellido_2008,Alnes_2006,Celerier2000}. However, unlike conventional local-void models, in which the dominant inhomogeneity is usually a radial variation in the matter density, the model considered here introduces a transition in the vacuum energy itself.
}

An apparent reduction in the dark energy density inferred from low-redshift observations could therefore be interpreted in several ways: as temporal evolution of the dark energy density, as spatial variation on scales comparable to the local cosmological environment, or as a combination of both. In the simplest models, these possibilities are closely related. A spatial transition between regions with different vacuum energies generically introduces a dynamical boundary, so the resulting configuration is spacetime-dependent.

A natural physical motivation for spatially varying dark energy is vacuum decay. If dark energy is associated with a scalar degree of freedom whose potential contains multiple local minima, the observed vacuum could correspond to a metastable false vacuum, with regions containing a lower-energy phase nucleating through quantum tunneling in a localized region of spacetime~\cite{Coleman_1977,callan_fate_1977,coleman_gravitational_1980}. The possibility of metastable vacua is familiar from attempts to understand inflation~\cite{guth_could_1983}, the cosmological constant problem~\cite{RevModPhys.61.1}, and the landscape of possible vacuum energies~\cite{Bousso:2000xa,Susskind:2003kw}, as well as from studies of electroweak vacuum metastability in the Standard Model (e.g., Refs.~\cite{Cabibbo:1979ay,Isidori:2001bm,Degrassi:2012ry,buttazzo_investigating_2013,markkanen_cosmological_2018}). Related ideas have also been explored in models where the present dark energy density is identified with a long-lived false vacuum (e.g., Refs.~\cite{Abbott:1984qf,Brown:1988kg,Feng_2001,Giddings:2003zw,Giddings:2004vr,Page:2006dt}), as we focus on in the present work. 

Because the rate of bubble nucleation is exponentially suppressed, it is most natural to expect vacuum decay to occur on a timescale either far shorter or far longer than the age of the Universe. Here, we assume that the timescale for decay is comparable to the age of the Universe and focus on spacetimes with a single bubble. We do not commit to a particular UV completion or microphysical tunneling mechanism that predicts this, although there could be theoretical motivations for considering what, at face value, looks like a finely tuned scenario. For example, a decay rate of order the Hubble time could prevent the unbounded growth of volume, e.g., eternal inflation (see e.g., Ref.~\cite{aguirre_eternal_2007} for a review), which has been conjectured to be incompatible with quantum gravity~\cite{Banks:2005ru,Aguirre:2006ap,Page:2006dt,Rudelius:2019cfh}. Alternatively, there could be selection effects making it more likely for observers to inhabit regions where decay happened only very recently. 

We take this picture as motivation for an effective geometric setup: a region in which the dark energy density is lower than in the surrounding spacetime nucleates and evolves dynamically since the vacuum energy difference across its boundary acts as a pressure difference, driving the wall outward. In the thin-wall limit, the boundary rapidly reaches ultra-relativistic speeds. Closely related systems have been studied in the context of bubbles of lower vacuum energy and cosmological phase transitions \cite{Sakai_1994, Pen_2014, Pen_1998, Aguirre_Matt_2005, Aguirre_Matt_2006, Johnson_2020, Berezin_1987, Sato_1986, Aurilia:1989sb, Blau_1987}. In particular, Sakai and Maeda \cite{Sakai_1994} analyzed the dynamics of an interface separating FLRW regions with different vacuum energies. In the minimal thin-wall construction, where two flat FLRW spacetimes with different effective cosmological constants are joined across a boundary, the Israel junction conditions lead to an equation of motion for the wall. Thus, while additional model-building could in principle produce an approximately static spatial profile, the simplest step-function-like transition in the dark energy density is expected to evolve in time.

In this work, we study this minimal bubble model as a phenomenological realization of spacetime varying dark energy. The model consists of a region of lower dark energy density embedded in a surrounding region of higher dark energy density, with the observer located inside the lower-density region. Within each region the dark energy is locally constant, with equation of state $w=-1$, so the model preserves the local structure of flat $\Lambda$CDM in each region. We maintain the boundary condition that the physical space must be continuous at the bubble wall while the scale factors, comoving radii, and time coordinates are free to have discontinuities. Additionally, we allow the observer to be displaced from the centre of the bubble, so that the model can generate anisotropic signatures. We then utilize observations to constrain the allowed observer location.

Although the thin-wall dynamics of such geometries have been studied extensively, their consequences for cosmological observables such as redshift and distance measures remain relatively unexplored. Previous work explored the observational imprints of continuous dark energy decay, e.g. Ref.~\cite{Shafieloo:2016bpk,Li:2019san}. More recently, Ref.~\cite{Bai:2026sux} explored the imprints of a completed phase transition on DESI BAO measurements. Closest to the current work, Ref.~\cite{Pen_2014} studied the imprint of a single dark energy bubble on the CMB, demonstrating the potential for strong constraints. Our goal is to determine how the single-bubble model for spacetime-varying dark energy modifies cosmological observables and to assess its viability in light of current data.

\begin{figure}
    \centering
    \includegraphics[width=0.9\linewidth]{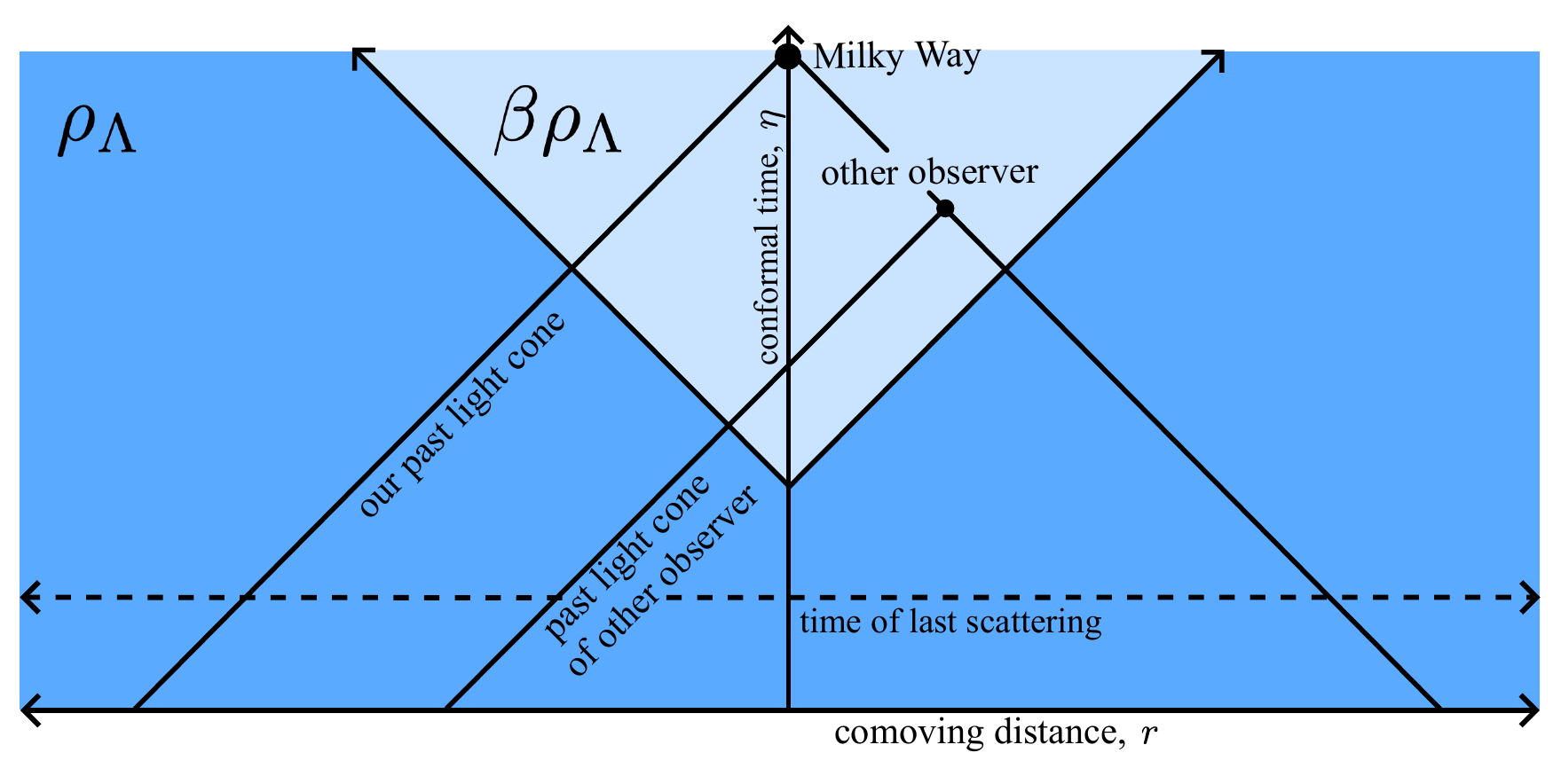}
    \caption{One-bubble cosmological spacetime. This is an edge-on view in 1+1 dimensions in comoving radius and the conformal time. The outer cosmology has a dark energy density of $\rho_\Lambda$ while the inner cosmology has a lower dark energy density of $\beta\rho_\Lambda$ where $0\leq\beta\leq1$. The wall moves radially outwards on an approximately null trajectory. If the Milky Way Galaxy lies approximately at the centre of the bubble, there is still observable anisotropic information carried to our galaxy from photons scattering off of non-central galaxies within the bubble. }
    \label{fig:ksz}
\end{figure}

\begin{figure}[htbp]
    \centering

    \begin{subfigure}{0.485\textwidth}
        \centering
        \includegraphics[width=\linewidth]{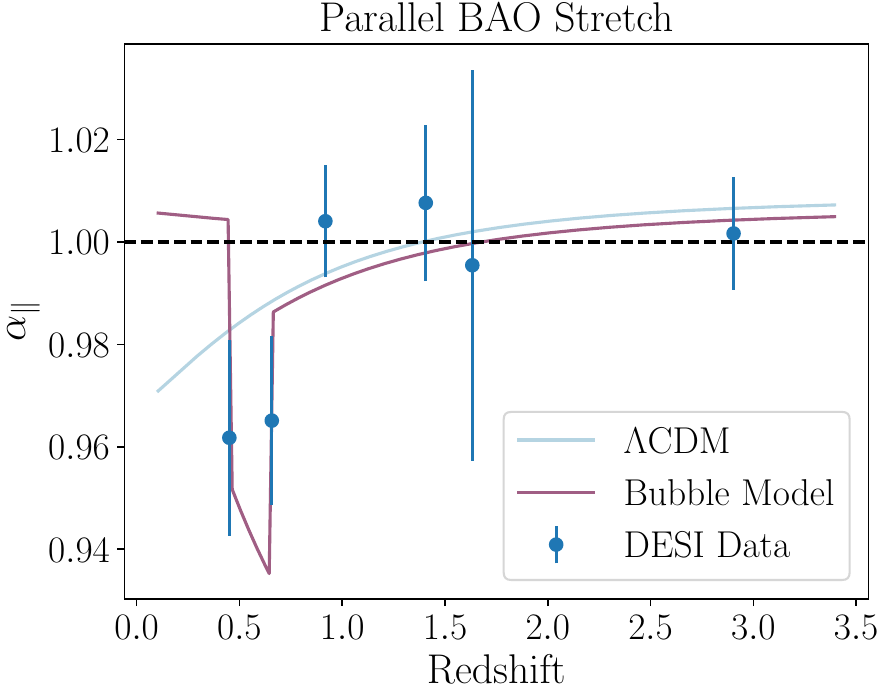}
        \label{fig:left-panel}
    \end{subfigure}
    \hspace{0.05cm}
    \begin{subfigure}{0.485\textwidth}
        \centering
        \includegraphics[width=\linewidth]{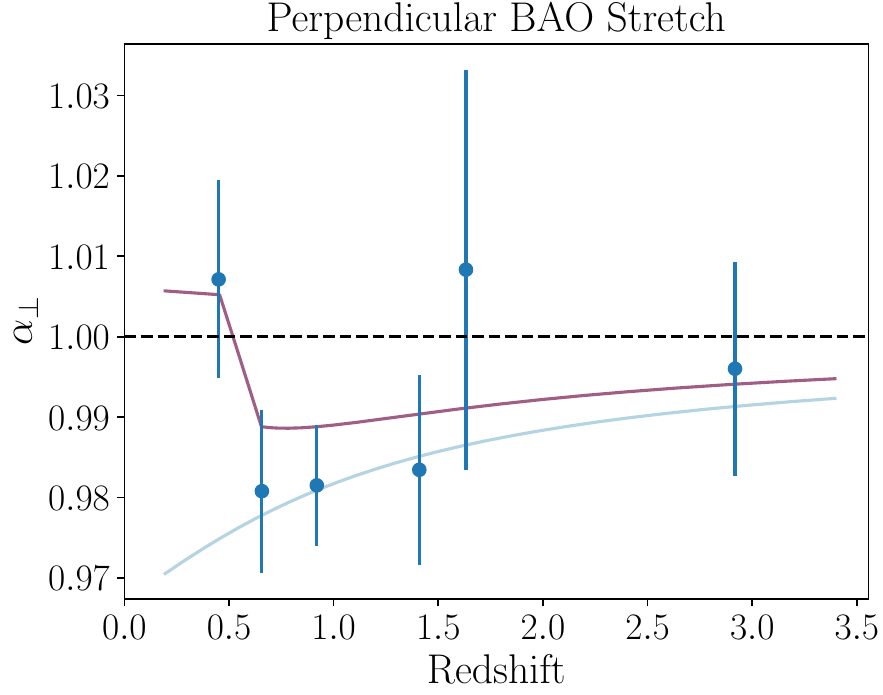}
        \label{fig:right-panel}
    \end{subfigure}

    \caption{Alcock--Paczynski Stretching parameters predicted by the Bubble Model for a bubble that nucleated at a redshift of 1.4 and that contains 10$\%$ less dark energy than the outer cosmology. The Model prediction is compared to both our fiducial $\Lambda$CDM cosmology and the DESI data.}
    \label{fig:two-panel}
\end{figure}

A schematic of the model is shown in Fig.~\ref{fig:ksz}. The bubble is 
characterized by its nucleation redshift, $z_{\rm nuc}$, and by the relative 
interior dark energy density $\beta\equiv\rho_{\Lambda,\mathrm{in}}/\rho_{\Lambda,\mathrm{out}}$. The bubble produces distinctive signatures in the Alcock--Paczynski BAO dilation parameters $\alpha_{\parallel}$ and $\alpha_{\perp}$, which quantify, respectively, the radial and transverse stretching of the BAO scale relative to a fiducial cosmology. As illustrated in Fig.~\ref{fig:two-panel}, suitable choices of $z_{\rm nuc}$ and $\beta$ generate sharp, localized features at low redshift that qualitatively resemble structure in the DESI measurements. Although this particular toy model is tightly constrained, these features demonstrate how an inhomogeneous late-time dark energy distribution can leave a signature in cosmological distance measurements.

The inhomogeneous geometry causes observers away from the centre of the bubble to experience a large CMB dipole. Scattering of this dipole by free electrons produces a kinetic Sunyaev--Zel'dovich (kSZ) signal that is observable from our location near the bubble centre. This scenario was first explored in Ref.~\cite{Pen_2014}. We use 
this and other CMB signals, along with the primary CMB and BAO measurements, to constrain the bubble parameters. We find that all CMB constraints plotted in Fig.~\ref{fig:CMB_combined_constraint} exclude, to 2$\sigma$, the region of parameter space that most closely reproduces the DESI-like BAO 
features.

The remainder of this paper is organized as follows. In Sec.~\ref{sec:bubble_model}, we develop the theoretical framework for computing cosmological observables in the bubble model. We first specify the cosmological parameters, geometry, assumptions, and bubble boundary matching prescription, and then derive the modifications to redshift, comoving, proper and angular-diameter distance, sound horizon scale, and Hubble Parameter. In Sec.~\ref{sec:obs}, we apply this framework to observables in three steps. First, we use the CMB dipole to constrain the location of the observer within the bubble, motivating the approximately central observer used in the rest of the data constraints and comparisons. Next, we compute the Alcock--Paczynski distortions induced by the bubble cosmology and compare the predicted parallel and perpendicular to the line-of-sight BAO stretching factors with DESI measurements. We then constrain the remaining bubble parameters — the nucleation redshift and the dark energy density difference between the interior and exterior regions — using additional CMB-derived observables such as the kSZ effect, the kSZ velocity-reconstruction monopole, and the distance to the last scattering surface. Since current supernova samples provide comparatively weak constraining power for the parameter space considered here, we present the corresponding luminosity-distance calculations and comparisons with Union3, Pantheon+, and Dark Energy Survey supernova data in Appendices \ref{append:luminosity_distance} and \ref{append:sne_comparison}.

\section{Bubble Model}
\label{sec:bubble_model}

We now define the model used throughout the calculation. The spacetime consists of two flat FLRW regions with different vacuum energy densities, joined across a thin spherical boundary. Since the geometry is homogeneous and isotropic only within each region, the usual FLRW distance--redshift relations cannot be applied globally without modification. We therefore begin by specifying the metric on each side of the wall and the matching conditions imposed at the boundary. These conditions determine the wall trajectory, which in turn fixes when photons cross from one cosmological region into the other.

After establishing the geometry and wall dynamics, we revisit the Friedmann equations inside the bubble. This step is important because the interior matter density and scale factor must be evolved consistently with the matching prescription at the bubble nucleation time. With this setup in place, we then derive the observables needed for comparison with data: redshift, comoving distance, proper and angular-diameter distances, the sound horizon scale, and the inferred Hubble parameter.

\begin{figure}
    \centering
    \includegraphics[width=0.8\linewidth]{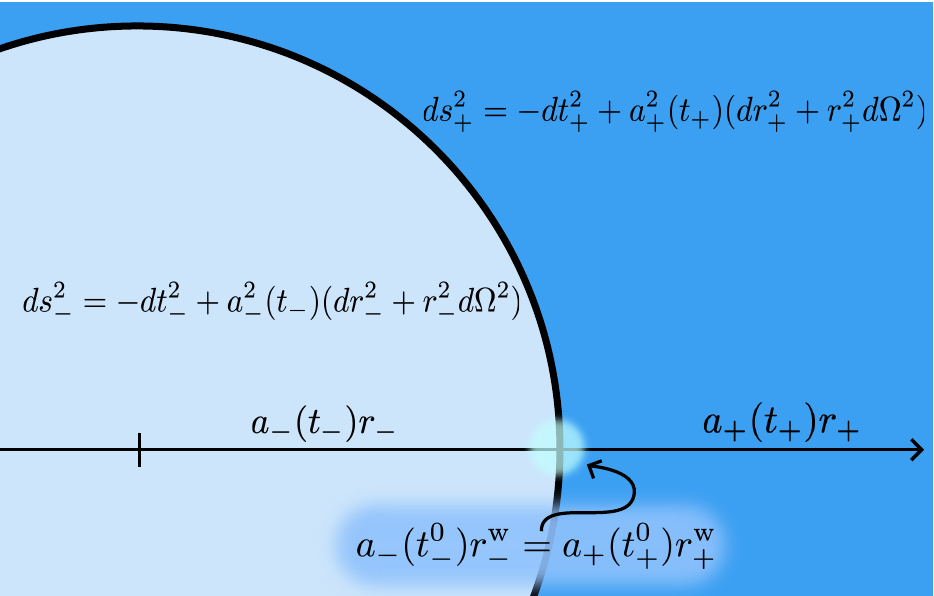}
    \caption{Illustration of the bubble at a constant time slice. The two glued FLRW spacetimes each have separate radial and time coordinates, but are related by the spatial continuity at the boundary of the bubble wall. Note that we use the thin-wall approximation where the spacetimes are directly glued together making the Christoffel symbols a step function at the wall and giving the stress-energy tensor a delta function contribution at the wall.}
    \label{fig:setup}
\end{figure}

\subsection{Metric and Junction Conditions}
\label{sec:metric_junction_cond}

As illustrated in Fig.~\ref{fig:setup}, we construct the bubble spacetime by joining two flat FLRW regions across a spherical thin wall. The interior region, denoted by a subscript $-$, has a smaller vacuum energy density $\rho_{\Lambda,-}$, while the exterior region, denoted by a subscript $+$, has a larger vacuum energy density $\rho_{\Lambda,+}$. As mentioned in Sec.~\ref{sec:intro}, the relationship between the two vacuum energy densities is given by $\rho_{\Lambda,-}  = \beta \rho_{\Lambda,+}$ where $0 \leq \beta\leq 1$. We allow the two regions to have separate scale factors, cosmic times, and comoving radial coordinates. The line elements on either side of the wall are
\begin{equation}
    ds^2_\pm = -dt_\pm^2 
    + a_\pm^2(t_\pm)\left(dr_\pm^2 + r_\pm^2 d\Omega^2\right),
    \qquad 
    d\Omega^2 = d\theta^2 + \sin^2\theta\,d\phi^2 .
    \label{eq:metric}
\end{equation}
Throughout this work we set $c=1$ unless explicitly stated.

The angular coordinates are identified across the wall by spherical symmetry. By contrast, the time coordinates $t_\pm$, comoving radial coordinates $r_\pm$, and scale factors $a_\pm(t_\pm)$ are defined separately in the two FLRW charts. This allows each side of the wall to retain the standard FLRW form, even though the spacetime as a whole is not globally FLRW.

The matching condition we impose is continuity of the physical, or areal, radius of the wall. If the wall is located at comoving radii $r_+(t_+)$ and $r_-(t_-)$ as described from the exterior and interior charts, respectively, then its physical radius must agree on the two sides:
\begin{equation}
    R(\tau) = a_+(t_+) r_+ = a_-(t_-) r_- ,
    \label{eq:boundary_relation}
\end{equation}
where $\tau$ denotes the proper time along the wall. Our setup follows that of Ref.~\cite{Sakai_1994}. This relation is the part of the junction condition most frequently used below: it relates the interior and exterior descriptions of the same physical boundary. Thus, while $t_\pm$, $r_\pm$, and $a_\pm$ need not be individually continuous across the wall, the combination $a_\pm(t_\pm)r_\pm$ is continuous at the boundary.

\subsection{Bubble Wall Trajectory}
\label{sec:wall_trajectory}

Before deriving the wall trajectory in the FLRW construction, it is useful to recall the basic intuition from vacuum decay~\cite{Coleman_1977}. In the thin-wall picture, a bubble of lower vacuum energy gains volume as it expands, while paying an energy cost in surface tension. Neglecting gravity and using a special-relativistic estimate, the energy released by the vacuum energy difference $\Delta \rho_\Lambda$ inside a bubble of physical radius $R$ is balanced against the relativistic energy of the wall according to
\begin{equation}
    \frac{4}{3}\pi R^3 \Delta\rho_\Lambda
    =
    \frac{4\pi R^2 \sigma}{\sqrt{1-v^2}},
\end{equation}
where $\sigma$ is the surface tension and $v$ is the velocity of the wall. This estimate suggests that once the bubble exceeds the critical radius
\begin{equation}
    R_c = \frac{3\sigma}{\Delta\rho_\Lambda},
\end{equation}
the wall rapidly accelerates toward the speed of light.

For the cosmological setup considered here, the wall motion can be derived more systematically from the Israel junction conditions, following the treatment of Ref.~\cite{Sakai_1994}. The first junction condition fixes the continuity of the areal radius across the wall, while the second junction condition relates the jump in extrinsic curvature to the wall stress-energy and determines the wall equation of motion. We give the full derivation for the present FLRW--FLRW matching in Appendix~\ref{append:bubble_wall_eom}.

The result is that, for the parameter range and cosmological length scales relevant to this work, the wall becomes ultra-relativistic on scales much smaller than those probed by the observables considered below. We therefore approximate the wall trajectory as an outgoing radial null trajectory in each FLRW region. In the coordinates of either side of the bubble, this gives
\begin{equation}
    \frac{d r_\pm}{dt_\pm} = \frac{1}{a_\pm(t_\pm)} ,
\end{equation}
where $r_\pm(t_\pm)$ denotes the comoving radius of the wall as described in the corresponding interior or exterior FLRW chart. We assume throughout that the bubble is centred at the origin and expands radially outward.

\subsection{Matter Content}
\label{sec:friedmann}

The matter density is taken to be continuous across the bubble wall at nucleation, and thereafter evolves separately in the two regions according to the corresponding local scale factor. Thus the interior and exterior matter densities agree at the transition time, but generally differ at later times when expressed at equal values of their respective cosmic times.

As discussed in Sec.~\ref{sec:metric_junction_cond}, we parametrize the vacuum energy density inside the bubble as a fraction of the exterior value,
\begin{equation}
    \rho_{\Lambda,-} = \beta \rho_{\Lambda,+},
    \qquad 0 \leq \beta \leq 1 ,
\end{equation}
where $\beta=1$ corresponds to no change in the vacuum energy and $\beta=0$ corresponds to an interior region with no vacuum energy component. The full derivation, including the relation between this parametrization and the usual cosmological variables, is given in Appendix~\ref{append:friedmann}. With the normalization chosen there, the scale factor for a flat matter-plus-vacuum FLRW region can be written as
\begin{equation}
    a(t) =
    \left(\frac{\Omega_{M,0}}{\beta\Omega_{\Lambda,0}}\right)^{1/3}
    \sinh^{2/3}\left(
    \frac{3}{2}\sqrt{\beta\Omega_{\Lambda,0}}\,H_0 t
    \right) .
    \label{eq:scale_factor}
\end{equation}
The exterior cosmology is recovered by setting $\beta=1$. The interior cosmology is described by the same expression with the reduced vacuum energy fraction $\beta\leq1$, together with the matter-density normalization fixed by equality of the matter density at nucleation.

The limiting case, $\beta=1$, removes the bubble since the vacuum energy is the same on both sides of the wall. The opposite limit, $\beta\rightarrow0$, gives a matter-dominated interior with no dark-energy component.

\subsection{Cosmological Parameters}
\label{sec:cosmo_param}

Throughout this work, the region exterior to the dark energy bubble is modelled as a spatially flat $\Lambda$CDM universe. For its late-time background evolution, we adopt
\begin{equation}
(h,\Omega_{M,0},\Omega_{\Lambda,0})
=(0.6736,0.2892,0.7108).
\end{equation}
Here, $\Omega_{M,0}=0.2892$ corresponds approximately to the lower $1\sigma$ limit of the DESI DR2 BAO constraint \cite{DESI_DR2}. This choice produces a somewhat better qualitative fit for our fiducial bubble model, allowing us to test whether it remains viable under assumptions that increase its compatibility with the low-redshift distance measurements. We adopt the Planck value of $h$, since DESI BAO alone does not determine $H_0$ independently, but instead constrains its degenerate combination with the sound horizon, $H_0r_d$. We also set the redshift at the surface of last scattering, $z_{\rm ls}$ to be 1089.92, from the Planck 2018 fit~\cite{Planck_2018_cosmo}.

Quantities determined by pre-recombination physics, most notably the sound horizon at the baryon-drag epoch, are calculated using the Planck physical densities
\begin{equation}
\omega_b\equiv\Omega_bh^2=0.02237,
\qquad
\omega_c\equiv\Omega_ch^2=0.1200,
\end{equation}
together with $N_{\rm eff}=3.046$ to determine $\Omega_{\nu,0}$ and $T_0=2.7255$ K, the temperature of the CMB today, measured by Ref.~\cite{Fixsen_2009} to determine $\Omega_{\gamma,0}$. The sound horizon is evaluated in the usual manner by integrating the ratio of the photon--baryon sound speed to the expansion rate over the pre-drag expansion history. At these redshifts, dark energy is negligible, while the expansion rate depends on the physical densities of baryons, cold dark matter, photons, and neutrinos. An accurate calculation therefore requires information beyond the total late-time matter fraction $\Omega_{M,0}$.

In particular, the DESI BAO fits do not separately determine the baryonic, cold-dark-matter, and neutrino contributions to $\Omega_{M,0}$. Attempting to construct the early-time expansion history directly from the DESI-preferred value of $\Omega_{M,0}$ would therefore require additional assumptions about how the total matter density should be divided among these components. Since we assume no modification to standard pre-recombination physics, we instead use the more precisely constrained and internally consistent Planck physical densities for the sound horizon calculation. These Planck parameters are used only to calibrate the early-time standard ruler and are not used to redefine the DESI-motivated late-time matter density adopted for the purposes of clearly constraining the bubble model with the most generous assumptions. 

\subsection{Distance Measures}
\label{sec:distance_measures}
We now derive the distance and expansion observables used to compare the bubble model with CMB and DESI measurements. In a globally FLRW cosmology, redshift, comoving distance, angular-diameter distance, the sound horizon, and the Hubble parameter are related by the standard homogeneous distance--redshift relations. In the bubble model these relations must be modified because a photon observed inside the bubble may have propagated through both the exterior and interior FLRW regions and crossed the moving wall along the way.

In the following subsections we compute the relevant observables in this piecewise-FLRW geometry: the observed redshift, the radial comoving distance, the proper and angular-diameter distances, the sound horizon scale, and the inferred Hubble parameter. These quantities enter the CMB and BAO constraints discussed in Sec.~\ref{sec:obs}. Intermediate derivations and algebraic details are collected in Appendices~\ref{append:friedmann}--\ref{append:z_theta}.

\subsubsection{Redshift}
\label{sec:redshift}
\begin{figure}
    \centering
    \includegraphics[width=0.75\linewidth]{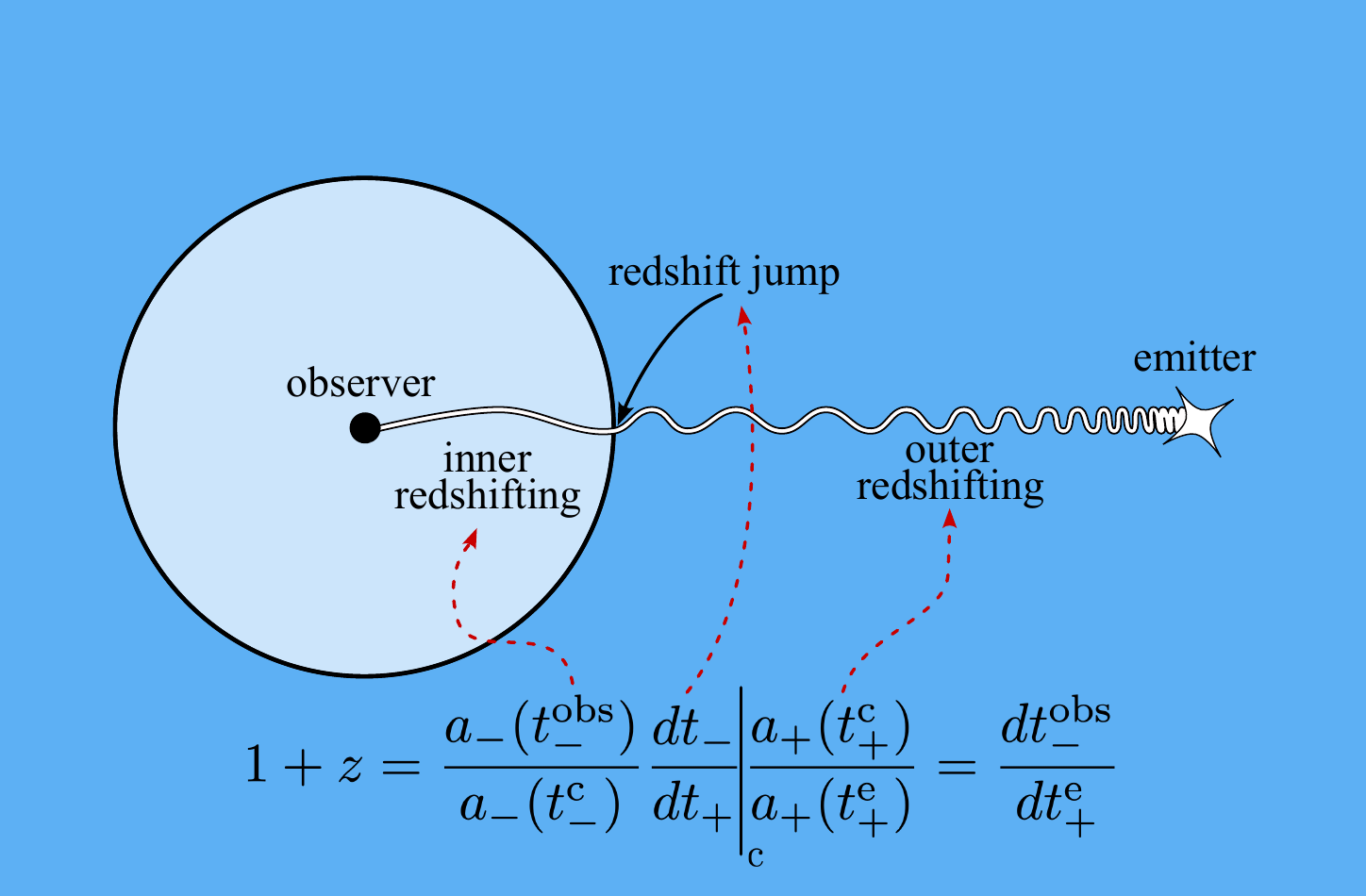}
    \caption{Illustration of bubble cosmology impact on photon redshifting. Note that less redshifting should take place within the bubble than without due to a lower cosmological constant within the bubble. Additionally, a redshift jump due to the coordinate change for respective comoving observers at the bubble wall produces a strong feature in the distance-redshift relationship.}
    \label{fig:photon_path}
\end{figure}

We first compute the redshift of a photon emitted in the exterior region and observed inside the bubble. Operationally, the redshift is obtained by comparing the separation between neighboring photon wavefronts at emission and observation,
\begin{equation}
    1+z = \frac{dt_-^{\rm obs}}{dt_+^{\rm e}} ,
    \label{eq:redshift}
\end{equation}
where \(dt_+^{\rm e}\) is the proper-time interval between wavefronts measured by a comoving emitter in the exterior cosmology, and \(dt_-^{\rm obs}\) is the corresponding interval measured by a comoving observer in the interior cosmology.

The same redshift can be written in the more physically transparent form
\begin{equation}
    1+z =
    \frac{a_-(t_-^{\rm obs})}{a_-(t_-^{\rm c})}
    \left.\frac{dt_-}{dt_+}\right|_{\rm c}
    \frac{a_+(t_+^{\rm c})}{a_+(t_+^{\rm e})}.
    \label{eq:redshift_product}
\end{equation}
Here \(t_\pm^{\rm c}\) denote the interior and exterior times assigned to the same wall-crossing event. The rightmost fraction is the usual FLRW redshift accumulated while the photon propagates in the exterior region, the leftmost fraction is the usual FLRW redshift accumulated after the photon enters the bubble, and the middle fraction accounts for the change between the exterior and interior cosmic-time coordinates at the wall. 

To evaluate the redshift numerically, we determine the map
\(
t_-^{\rm obs}(t_+^{\rm e})
\)
by following the photon path from emission, through the wall-crossing event, to observation, as illustrated in Fig.~\ref{fig:photon_path}. Within each FLRW region, radial null geodesics satisfy
\begin{equation}
    \frac{dr_{\pm}^{\gamma}}{dt_\pm}
    =
    \pm \frac{1}{a_\pm(t_\pm)} ,
    \label{eq:photon_null_geodesic}
\end{equation}
with the sign chosen according to the direction of propagation. In our convention, for a photon traveling inward toward the observer situated at the origin, we choose the minus sign. The bubble wall, approximated as an outgoing radial null trajectory, obeys
\begin{equation}
    \frac{dr_{\pm}^{\rm w}}{dt_\pm}
    =
    \frac{1}{a_\pm(t_\pm)} .
    \label{eq:wall_null_trajectory}
\end{equation}

Starting from an emission event \((t_+^{\rm e},r_+^{\rm e})\), we evolve the exterior photon trajectory \(r_{+}^\gamma(t_+)\) and the exterior wall trajectory \(r_{+}^{\rm w}(t_+)\). The crossing event is found by solving
\begin{equation}
    r_{+}^\gamma(t_+^{\rm c})
    =
    r_{+}^{\rm w}(t_+^{\rm c}) .
\end{equation}
This determines \(t_+^{\rm c}\) and \(r_+^{\rm c}\). We then use the wall matching condition,
\begin{equation}
    a_+(t_+^{\rm c}) r_+^{\rm w}
    =
    a_-(t_-^{\rm c}) r_-^{\rm w},
\end{equation}
to numerically determined wall mapping \(t_-(t_+)\) and obtain the corresponding interior crossing event \((t_-^{\rm c},r_-^{\rm c})\). Finally, we evolve the interior photon trajectory until it reaches the observer at \(r_-^{\rm obs}\), giving \(t_-^{\rm obs}\). Repeating this procedure for neighboring emission times gives \(t_-^{\rm obs}(t_+^{\rm e})\), and hence the redshift through Eq.~\eqref{eq:redshift}.

The present interior time \(t_-^0\) provides one remaining normalization of the light cone. We fix this by requiring that the observed redshift of the last-scattering surface be \(z_{\rm ls}=1089.92\). Since the bubble nucleates at low redshift in the scenarios considered here, the microphysics determining recombination is unchanged, and the exterior time of last scattering \(t_{+}^{\rm ls}\) can be computed from the exterior Friedmann solution in the usual way. Requiring that this event be observed today with redshift \(z_{\rm ls}\) determines \(t_-^{0}\). With this normalization fixed, we can compute the redshift--distance relation for sources observed inside the bubble. Fig.~\ref{fig:r_vs_z} shows an example of the resulting redshift--comoving-distance relation.

Throughout this work we label the bubble nucleation time by the redshift it would correspond to in the exterior cosmology. More precisely, we define
\[
    1+z_{\rm nuc} \equiv \frac{1}{a_+(t_{+}^{\rm nuc})},
\]
where the exterior scale factor is normalized such that \(a_+(t_{+}^0)=1\) and the superscript, 0, refers to the time associated with the outside cosmological parameters, $\Omega_{M,0}$ and $\Omega_{\Lambda,0}$. This quantity should not be interpreted as the redshift measured by an observer inside the bubble. Rather, it is a convenient parametrization of the nucleation time with respect to the homogeneous exterior \(\Lambda\)CDM background. With this convention, strictly in the outside cosmology, the usual intuition for cosmological redshift applies: \(z=0\) corresponds to the present epoch of the exterior cosmology, \(z\simeq 0.3\) to the onset of dark-energy domination, and \(z\simeq 1089.92\) to last scattering.

\subsubsection{Comoving, Proper, and Angular-Diameter Distances}
\label{sec:co-move_proper_ang_diam}
\begin{figure}
    \centering
    \includegraphics[width=\linewidth]{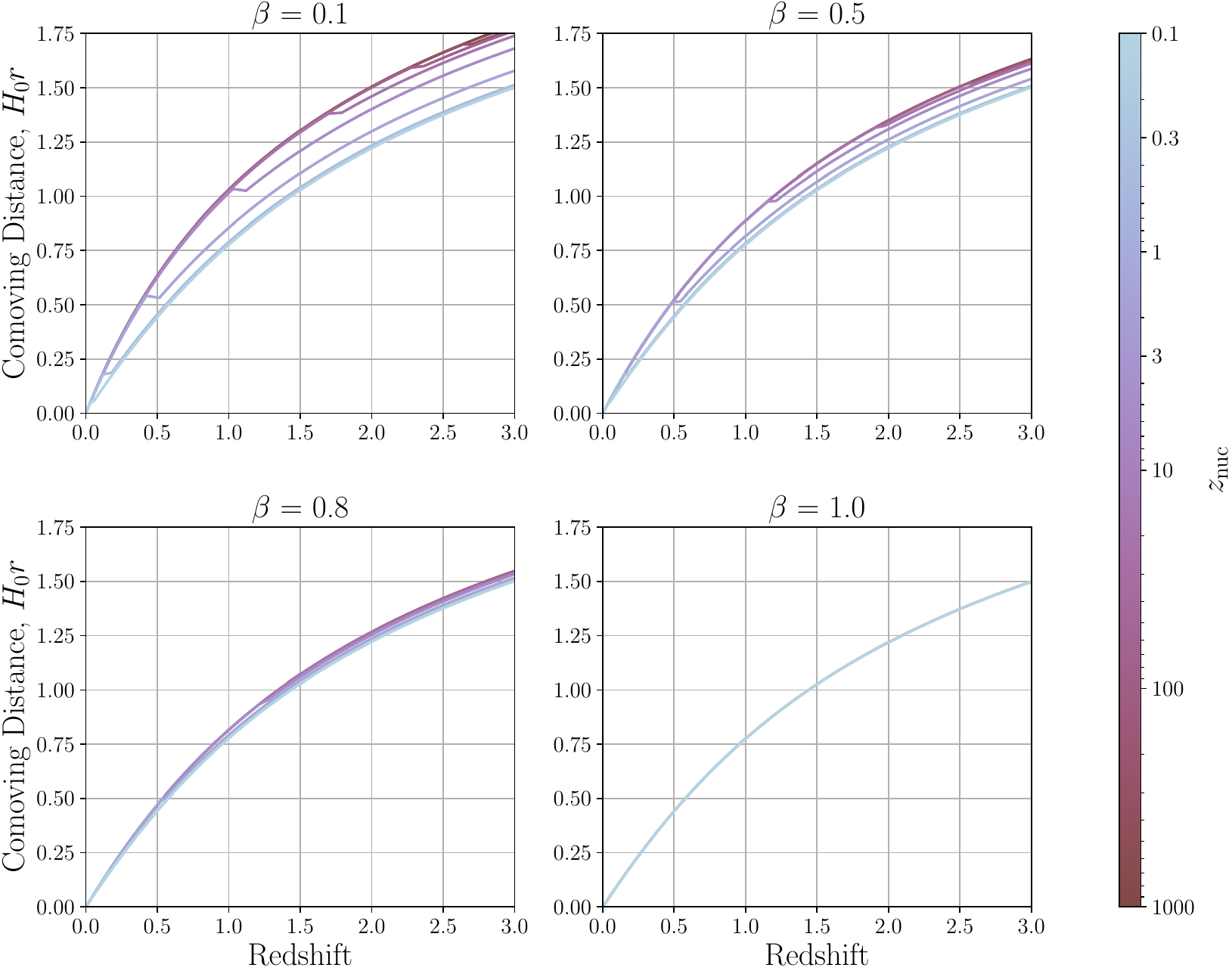}
    \caption{Comoving distance-redshift relationship in a bubble cosmology. The relationship for varying $\beta$ and $z_{\rm nuc}$ values is plotted. A discontinuity occurs from a jump in redshift at the bubble wall.}
    \label{fig:r_vs_z}
\end{figure}

Once the redshift relation has been determined, the comoving radius of the emission event can be obtained numerically by inverting the function \(z(r^{\rm e})\). This gives the comoving distance to the source as a function of the observed redshift for the chosen source--observer configuration, as shown in Fig. \ref{fig:r_vs_z}. 

For an observer located at the origin, the proper radial distance to the source at the observation time is simply the physical radius of the source measured from the origin,
\begin{equation}
    D_{\rm p}(t_\pm^{\rm obs}) = a_\pm(t_\pm^{\rm obs})\, r_\pm^{\rm e},
\end{equation}
where the \(+\) or \(-\) description may be used according to which FLRW chart contains the source. Because the two charts describe the same physical spacetime and the areal radius is continuous across the wall, this quantity is well defined provided the corresponding interior and exterior observation times are related using the wall time map \(t_-(t_+)\). Although this expression has the same form as in a homogeneous FLRW cosmology, the resulting distance generally differs from the standard case because fixing the observed redshift of the CMB also fixes a different present interior time \(t_-^{0}\), so in general \(a_-(t_-^{0})\neq 1\).

For an observer within the bubble but away from the origin, the radial proper distance to a source lying along the same line of sight can be written as the difference of the corresponding physical radii,
\begin{equation}
    D_{\rm p}
    =
    \left| a_+(t_+^{\rm obs})\,r_+^{\rm e}
    -
    a_-(t_-^{\rm obs})\,r_-^{\rm obs}\right| ,
\end{equation}
where \(r_-^{\rm obs}\) is the observer position in the interior chart and \(r_+^{\rm e}\) is the source position in the exterior chart. This expression assumes that the source and observer lie on the same radial line; more general configurations would require a more general geometric treatment; see Appendix~\ref{append:snell}.

We now turn to the angular-diameter distance. For a source of transverse physical size \(\ell_\perp\) subtending a small angle \(\theta\), the angular-diameter distance, both in homogeneous FLRW and the bubble model, is defined by
\begin{equation}
    D_A \equiv \frac{\ell_\perp}{\theta}.
    \label{eq:angular_diameter_distance}
\end{equation}

\begin{figure}
    \centering
    \includegraphics[width=\linewidth]{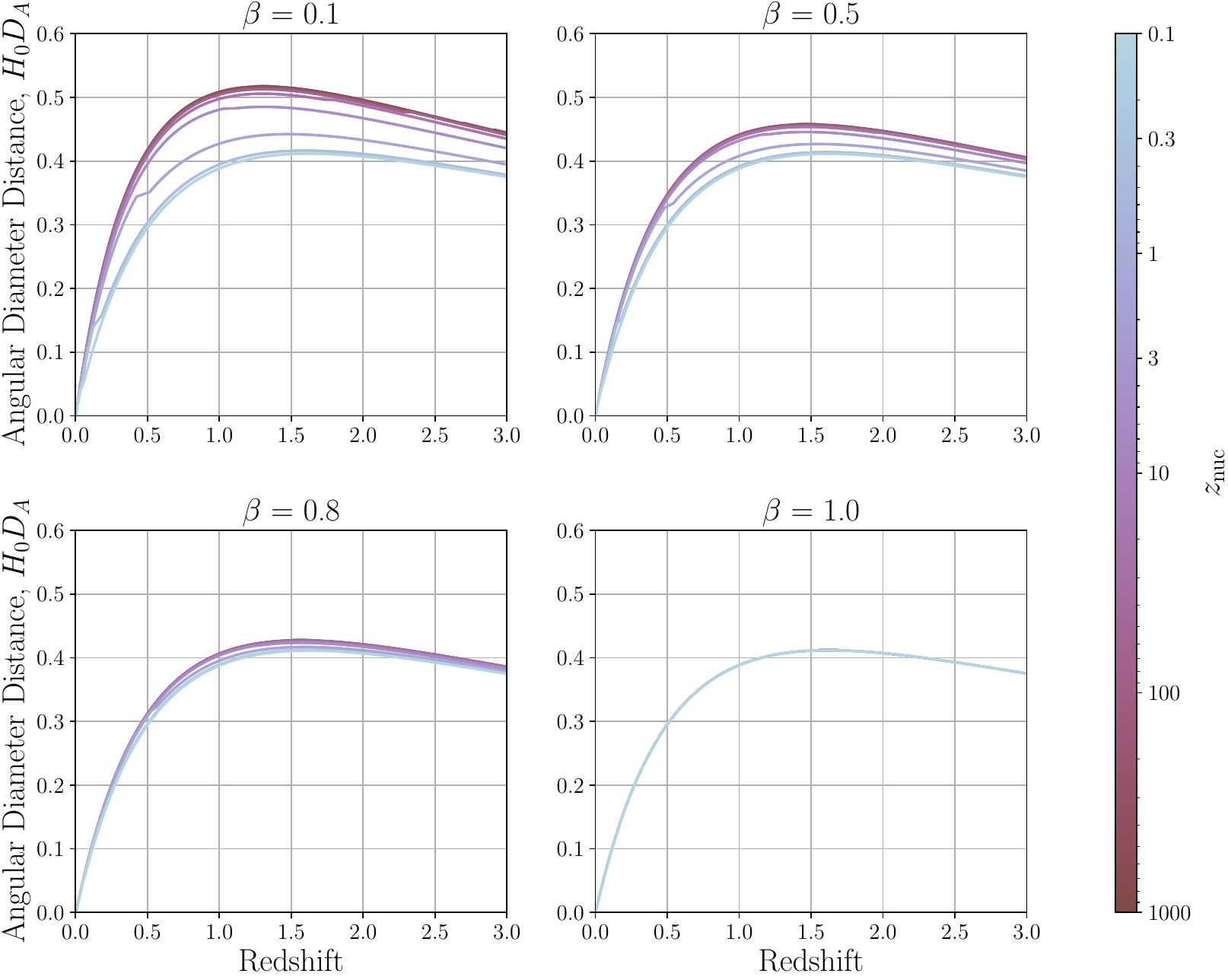}
    \caption{Angular-diameter distance-redshift relationship in a bubble cosmology. The relationship for varying $\beta$ and $z_{\rm nuc}$ values is plotted. A discontinuity occurs from a jump in redshift at the bubble wall.}
    \label{fig:DA}
\end{figure}

The nontrivial point in the bubble spacetime is that, for a central observer and radially propagating photons, the angular-diameter distance retains the same geometric form as in a homogeneous flat FLRW universe:
\begin{equation}
    D_A
    =
    \frac{\ell_\perp}{\theta}
    =
    a_\pm(t_\pm^{\rm e})\,r_\pm^{\rm e}.
    \label{eq:angular_diameter_bubble}
\end{equation}
Here \(r_\pm^{\rm e}\) is the radial coordinate of the emission event in the FLRW chart containing the source. Although the photon may cross the bubble wall before reaching the observer, the spherical symmetry about the central observer preserves the usual small-angle relation between the observed angle and the transverse physical size at emission. This is the result derived explicitly in Appendix~\ref{append:angular_diameter_dist_calc}. A plot of $D_A$ with various nucleation redshifts and values of $\beta$ can be seen in Fig. \ref{fig:DA}.

Thus \(D_A\) and \(D_{\rm p}\) are closely related in form, but they are not the same quantity: \(D_A\) is evaluated at emission, while \(D_{\rm p}\) is a evaluated at observation. 
For an observer at the centre of the bubble, the geometry remains isotropic and \(D_A\) depends only on redshift. For an off-centre observer the bubble introduces a preferred direction, and the relation between angular size and physical transverse size becomes direction dependent. This anisotropy will be relevant when discussing observables such as the sound horizon scale in Sec.~\ref{sec:sound_horizon}.

\subsubsection{Sound Horizon Size}
\label{sec:sound_horizon}

The sound horizon provides the standard ruler used by both CMB and BAO measurements. In a homogeneous cosmology, the comoving sound horizon \(r_s\) can be computed from the early-universe density parameters, including the photon density \(\Omega_{\gamma,0}\), which is fixed by the measured CMB temperature today~\cite{Fixsen_2009}, and the baryon density \(\Omega_{b,0}\), which is measured by Planck to roughly \(0.3\%\) precision~\cite{Planck_2018_cosmo}. However, the CMB acoustic angular scale \(\theta_s\) is measured significantly more precisely, at roughly the \(0.03\%\) level~\cite{Planck_2018_cosmo}. We therefore choose to determine \(r_s\) using the observed angular scale \(\theta_s\) together with the bubble-model distance to last scattering, rather than computing \(r_s\) directly from the early-universe density parameters. Note that we use the $0.3\%$ margin of error for $r_s$ from Planck as a CMB constraint on the distance to the surface of last scattering, outlined in Sec.~\ref{sec:sound_horizon_constraints}.

For a central observer, the angular-diameter distance relation derived in Appendix~\ref{append:angular_diameter_dist_calc} gives
\begin{equation}
    \theta_s
    =
    \frac{a_+(t_{+}^{\rm ls})\,r_s}
         {a_+(t_{+}^{\rm ls})\,r_+^{\rm ls}}
    =
    \frac{r_s}{r_+^{\rm ls}},
    \label{eq:theta_s_cancellation}
\end{equation}
or equivalently
\begin{equation}
    r_s = \theta_s r_+^{\rm ls}.
    \label{eq:rs_from_theta}
\end{equation}
Here \(r_s\) is the comoving sound horizon at last scattering, \(r_+^{\rm ls}\) is the exterior transverse comoving distance to the last-scattering surface, and \(t_{+}^{\rm ls}\) is the exterior time of last scattering. The scale factors cancel because both the transverse physical sound horizon size and the angular-diameter distance to last scattering are evaluated at the same emission time.

The distance \(r_+^{\rm ls}\) is computed using the bubble-model relationship between comoving distance and observed redshift, together with the assumption that the observed redshift to the surface of last scattering is $z_{\rm ls} = 1089.92$. Since the phase transition occurs only at late times in the scenarios considered here, the microphysics determining recombination is unchanged, and the value of \(z_{\rm ls}\) may be taken to be the standard one. We take \(\theta_s\) directly from Planck~\cite{Planck_2018_cosmo}, and use Eq.~\eqref{eq:rs_from_theta} to infer the value of \(r_s\) in the bubble model. 

\begin{figure}
    \centering
    \includegraphics[width=\linewidth]{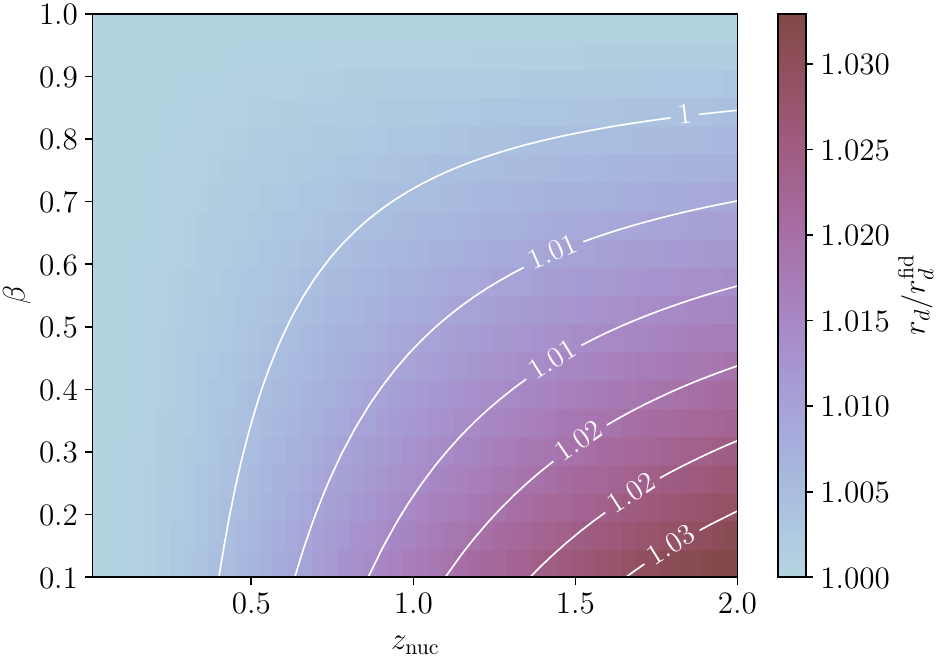}
    \caption{
    Comoving drag radius, \(r^d\), as a function of bubble-model parameters, normalized by the fiducial $r_d^{\mathrm{fid}}$ computed with Planck 2018 cosmological parameters \cite{Planck_2018_cosmo}.
    }
    \label{fig:rd_heatmap}
\end{figure}

For BAO observables, the relevant standard ruler is the sound horizon at the baryon drag epoch, \(r_d\), rather than the sound horizon at photon decoupling, \(r_s\). The drag epoch occurs at a slightly lower redshift than last scattering, with \(z_d\simeq1060\) compared to \(z_{\rm ls}\simeq1090\). We therefore obtain \(r_d\) from \(r_s\) by integrating over the relatively small redshift interval between \(z_{\rm ls}\) and \(z_d\). Computing this integral of $c_s(z)/H(z)$ over this range of redshift requires information about the radiation contribution; we therefore use the Planck 2018 values as mentioned in Sec.~\ref{sec:cosmo_param}. The model values of $r_d$ over our parameter space are plotted as a heatmap in Fig.~\ref{fig:rd_heatmap}. We use the resulting value of $r_d$ for our model predictions of the Alcock--Paczynski stretch parameters in Sec.~\ref{sec:par_bao} and Sec.~\ref{sec:perp_bao}.

After the bubble nucleates, a BAO shell that was initially comoving with the exterior cosmology may be swept into the bubble. Its physical size is continuous across the wall, but its comoving size changes because the interior and exterior scale factors differ. Applying the boundary relation to the physical BAO scale gives
\begin{equation}
    a_+(t_+^{\rm c})\, r^d_+
    =
    a_-(t_-^{\rm c})\, r^d_- ,
    \label{eq:rd_boundary_relation}
\end{equation}
so that the drag-epoch ruler expressed in interior comoving coordinates is
\begin{equation}
    r^d_-(z_{\rm obs})
    =
    \frac{
    a_+\!\left(t_+^{\rm c}(z_{\rm obs})\right)
    }{
    a_-\!\left(t_-^{\rm c}(z_{\rm obs})\right)
    }
    r^d_+ .
    \label{eq:rd_inner}
\end{equation}
Here \(r^d_+\equiv r_d\) is the usual exterior comoving sound horizon at the drag epoch, while \(r^d_-(z_{\rm obs})\) is the same physical ruler expressed in the interior comoving coordinates after crossing the wall. The dependence on \(z_{\rm obs}\) arises because shells observed at different redshifts cross the bubble wall at different times. They therefore enter the interior cosmology with different ratios of exterior to interior scale factor. The computed difference between $r^d_+$ and $r^d_-$ is negligible and therefore $r^d_+$ is used.

\subsubsection{Hubble Parameter}
\label{sec:hubble}
\begin{figure}
    \centering
    \includegraphics[width=0.9\linewidth]{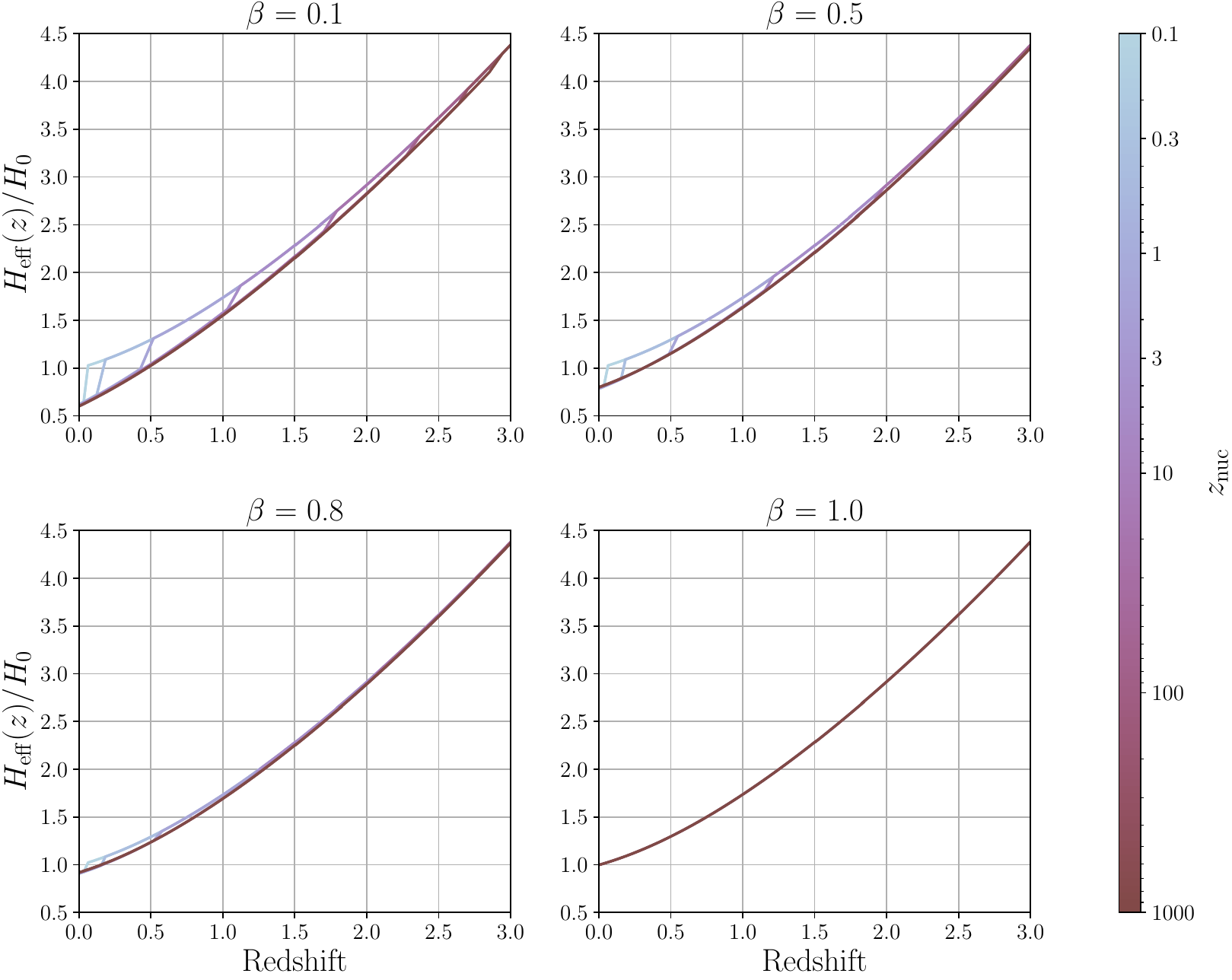}
    \caption{Hubble parameter as a function of redshift in a bubble cosmology. The relationship for varying $\beta$ and $z_{\rm nuc}$ values is plotted. A discontinuity occurs from a jump in redshift at the bubble wall. Note that $H_{\rm eff}(z)$ tracks an inner and an outer trajectory for each $\beta$ value that is independent of $z_{\rm nuc}$ (aside from governing crossing time over the past light cone), as shown in Appendix \ref{append:friedmann}.}
    \label{fig:hubble}
\end{figure}

The Hubble parameter enters BAO measurements through the line-of-sight distance scale. In a homogeneous FLRW cosmology, the radial comoving distance is
\begin{equation}
    r(z) = \int_0^z \frac{dz'}{H(z')} ,
\end{equation}
so that
\begin{equation}
    \frac{dr}{dz} = \frac{1}{H(z)} .
\end{equation}
Equivalently, the line-of-sight Hubble distance is
\begin{equation}
    D_H(z) \equiv \frac{1}{H(z)}
    =
    \frac{dr}{dz}.
\end{equation}

In the bubble model, the spacetime is not globally FLRW, so \(H(z)\) should be interpreted as the value that would be inferred from the observed radial distance--redshift relation. We therefore define an effective line-of-sight Hubble parameter by
\begin{equation}
    H_{\rm eff}(z)
    \equiv
    \left(\frac{dr}{dz}\right)^{-1},
    \label{eq:hubble_eff}
\end{equation}
where \(r(z)\) is the comoving distance inferred using the bubble-model light cone.

This definition makes clear why the bubble wall can produce a sharp feature in the inferred Hubble parameter. Near the wall-crossing redshift, a small change in comoving emission radius can correspond to a relatively large change in observed redshift. Equivalently, \(dr/dz\) becomes small, and the effective \(H_{\rm eff}(z)\) develops a step-like feature as seen in Fig.~\ref{fig:hubble}. This feature should be understood as a consequence of the modified distance--redshift relation across the wall, rather than as a local divergence of the expansion rate within either FLRW region.

\section{Observational Signatures and Constraints}
\label{sec:obs}
The bubble model contains three phenomenological parameters: the observer's position within the bubble, the nucleation redshift of the bubble ($z_{\rm nuc}$), and the fractional interior dark energy density ($\beta$). The observer position determines the degree of anisotropy seen in the sky, while the nucleation redshift controls the present size of the bubble and the redshift at which the wall intersects the past light cone. The parameter \(\beta\) controls the difference between the interior and exterior expansion histories.

We first use the CMB dipole measured by Planck~\cite{Planck_2018_cosmo} to constrain the observer's location within the bubble. We then compare the predicted BAO Alcock--Paczynski (AP) distortions with the DESI measurements. We find that the bubble model produces suggestive AP features for some choices of parameters. However, we then consider additional constraints from the kinematic Sunyaev--Zel'dovich (kSZ) effect, the kSZ velocity-reconstruction monopole, and the sound horizon calibration, which together restrict the allowed nucleation redshift and interior dark-energy fraction. The CMB constraints only allow bubbles at values of $z_{\rm nuc}$ below those that can be probed by DESI.

\subsection{Observer-position Constraint from the CMB Dipole}
\label{sec:cmb_dipole_constraints}

\begin{figure}
    \centering
    \includegraphics[width=0.65\linewidth]{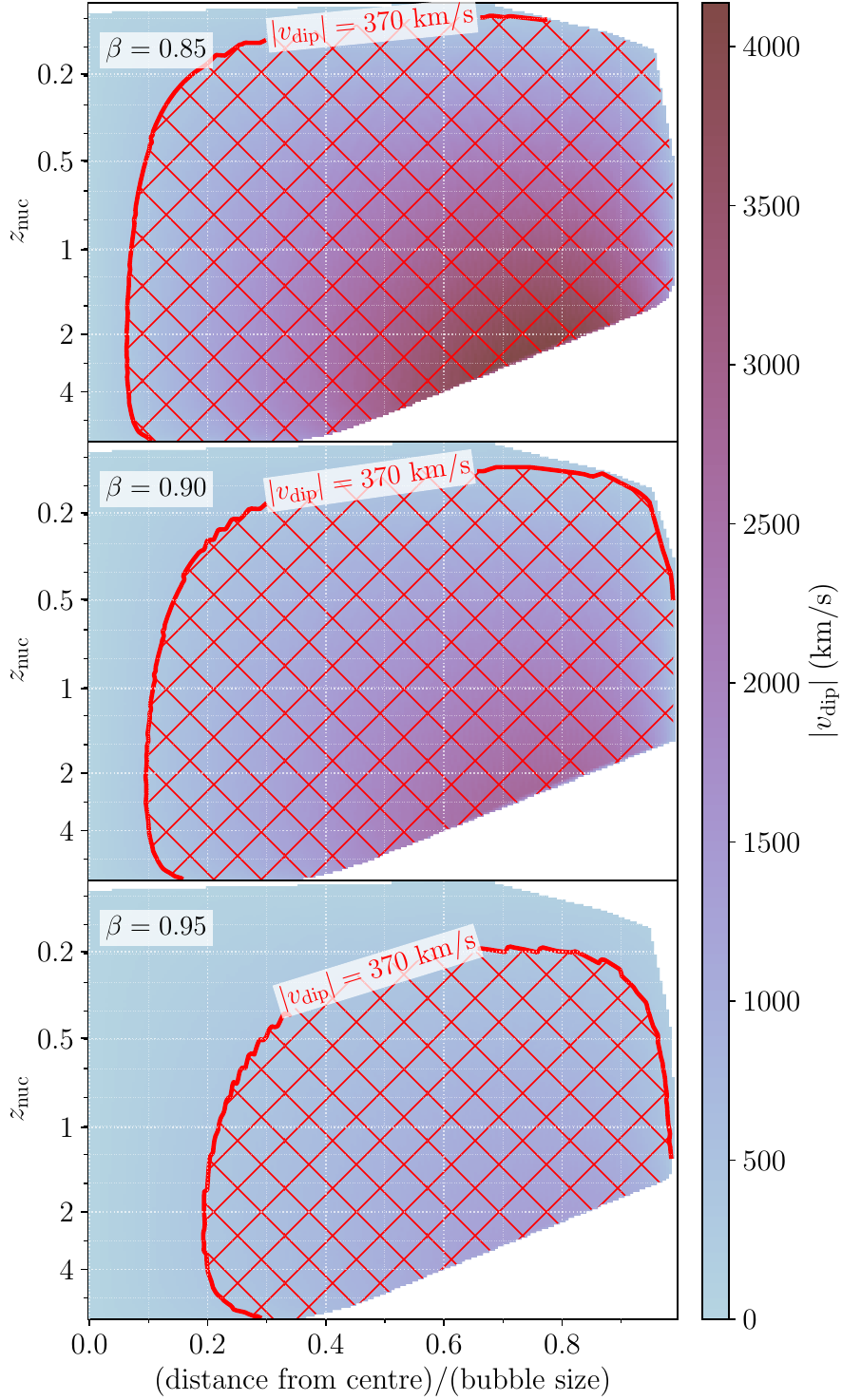}
    \caption{The bubble contribution to the CMB dipole plotted for varying values of $z_{\rm nuc}$, $\beta$, and observer distance to the bubble centre normalized by the bubble size at the time of observation. The dipole is converted to a velocity to compare to the CMB bound of $~370$km/s \cite{Planck_2018_dipole}. Observers inside the bubble are constrained to be in the parameter space plotted to be in the blue region shown without cross hatching. The region of the parameter space plotted with cross hatching is ruled out. }
    \label{fig:dipole_constraint}
\end{figure}

For an observer located at the centre of the bubble, the spacetime remains isotropic about the observer, and the CMB redshift is independent of direction. For an off-centre observer, however, radial symmetry about the observer is broken. Photons arriving from different directions cross the bubble wall at different times, and therefore experience different redshift histories. As a result, the observed CMB redshift becomes angle dependent,
\begin{equation}
    z_{CMB} = z(\theta),
\end{equation}
where \(\theta\) is the angle between the line of sight and the direction from the bubble centre to the observer. The geometric construction used to compute \(z(\theta)\) is described in Appendix~\ref{append:z_theta}.

The angular dependence of \(z(\theta)\) induces an apparent CMB temperature dipole. Since the observed temperature scales as \(T(\theta)\propto [1+z(\theta)]^{-1}\), the fractional temperature variation relative to the mean redshift \(z^{\rm avg}\) is
\begin{equation}
    \frac{\Delta T(\theta)}{T}
    =
    \frac{1+z^{\rm avg}}{1+z(\theta)} - 1  = \frac{z^{\rm avg}-z(\theta)}{1+z(\theta)}.
    \label{eq:cmb_temp_anisotropy_from_ztheta}
\end{equation}

We extract the dipole component by projecting this angular dependence onto the first Legendre polynomial. Equivalently, the apparent velocity dipole is
\begin{equation}
    v_{\rm dip}
    =
    \frac{3}{2}
    \int_0^\pi
    \left(
    \frac{z^{\rm avg}-z(\theta)}{1+z(\theta)}
    \right)
    \sin\theta \cos\theta\, d\theta .
    \label{eq:velocity_dipole}
\end{equation}
Here we have set \(c=1\), so \(v_{\rm dip}\) is dimensionless. In practice, we take
\begin{equation}
    z^{\rm avg} = 1089.92
\end{equation}
as the mean CMB redshift~\cite{Planck_2018_cosmo}, and require the induced dipole not to exceed the CMB dipole measured by Planck~\cite{Planck_2018_dipole}. This gives an upper bound on the observer's displacement from the bubble centre for each choice of the bubble parameters.

Fig. \ref{fig:dipole_constraint} shows the resulting constraint on observer position for several values of \(\beta\). The allowed region lies outside of the contours, corresponding to observers sufficiently close to the bubble centre that the induced redshift dipole remains below the observed CMB dipole. White regions indicate parameter combinations for which CMB photons do not intersect the bubble wall. This happens for three possible reasons: either the bubble nucleated early enough that the CMB photons reaching us today initially scattered while inside the bubble, the bubble nucleated later than the CMB photons would have reached our location in space, or the bubble geometry alters the distance to the surface of last scattering enough that the CMB photons would never reach the bubble wall. 

Although some additional freedom appears near the bubble wall for weak dark-energy contrast, see \(\beta=0.95\) in the bottom panel of Fig.~\ref{fig:dipole_constraint}, the observer is generally constrained to lie close to the centre of the bubble. This additional allowed region corresponds to cases in which the wall has only recently passed the observer. In this scenario, only a small portion of the
observer's past light cone intersects the region affected by the bubble.
Consequently, the bubble modifies the CMB redshift over only a small angular
region of the sky, while most lines of sight remain effectively indistinguishable from fully those in the exterior $\Lambda$CDM cosmology. The
bubble would therefore have little effect on observables averaged over the
full sky. Therefore, in the remainder of
the paper, we focus on the simplified case of an approximately central observer.

\subsection{Comparison with DESI BAO Features}

\label{sec:DESI_BAO_compare}

We now compare the qualitative features of the bubble model with the BAO distance measurements reported by DESI.

For the exterior cosmology used in computing the predicted parallel and perpendicular BAO stretch parameters, we adopt the DESI best-fit cosmological parameters. The DESI-preferred matter density is lower than the Planck best-fit value, and this lower \(\Omega_{M,0}\) has a noticeable effect on the predicted BAO stretches. For the illustrative comparison below, we use the lower \(1\sigma\) value of the DESI-inferred \(\Omega_{M,0}\), which gives a good match to the observed trend. Other choices within the allowed DESI parameter range can be explored, but we find that the stretch parameters are more sensitive to \(\Omega_{M,0}\) than to the dimensionless Hubble parameter \(h\).

The comparison in this subsection is intended as a feature-level comparison to motivate future studies. Importantly, we do not vary cosmological parameters or provide a quantitative assessment of the goodness of fit to this extended model. We will see in Sec. \ref{sec:additional_cmb}, the CMB constrains the bubble model to regions of the parameter space where it is indistinguishable from our fiducial $\Lambda$CDM predictions.

\subsubsection{Parallel BAO Stretch}
\label{sec:par_bao}

\begin{figure}
    \centering
    \includegraphics[width=\linewidth]{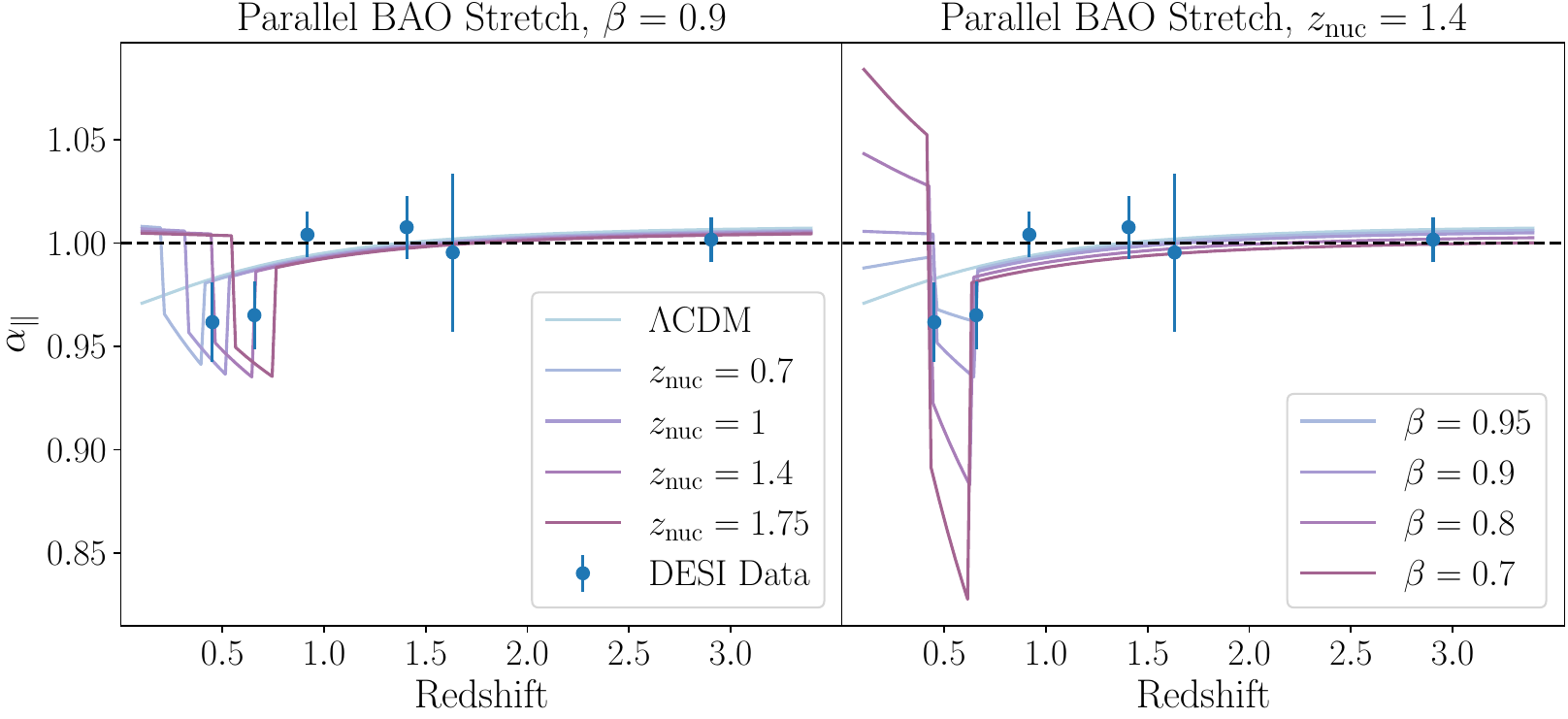}
    \caption{The Alcock--Paczynski parallel stretch parameters are plotted with varying $\beta$ and $z_{\rm nuc}$ values.}
    \label{fig:par_stretch}
\end{figure}

We compute the parallel BAO stretch parameter, \(\alpha_{\parallel}\), in the bubble model and compare it with the DESI BAO measurements. The parallel stretch quantifies the change in the inferred BAO scale along the line of sight relative to the fiducial cosmology used in the BAO analysis. Following the standard Alcock--Paczynski stretch convention~\cite{Alam_2017,Beutler_2011,Alam_2021, DESI_DR1, DESI_DR2}, we define
\begin{equation}
    \alpha_{\parallel}(z_{\rm obs})
    =
    \frac{
    H^{\rm fid}(z_{\rm obs})\, r_d^{\rm fid}
    }{
    H_{\rm eff}^{\rm model}(z_{\rm obs})\, r_d^{\rm model}
    } .
    \label{eq:alpha_parallel}
\end{equation}
Here \(H_{\rm eff}^{\rm model}(z_{\rm obs})\) is the effective line-of-sight Hubble parameter inferred from the bubble-model distance--redshift relation,
\begin{equation}
    H_{\rm eff}^{\rm model}(z_{\rm obs})
    =
    \left(\frac{dr}{dz_{\rm obs}}\right)^{-1},
\end{equation}
as described in Sec.~\ref{sec:hubble}. The quantity \(r_d^{\rm model}(z_{\rm obs})\) is the drag-epoch sound horizon expressed in the appropriate comoving coordinates for the observed shell, as described in Sec.~\ref{sec:sound_horizon}.

For the fiducial quantities, we use the Planck 2018 cosmological parameters~\cite{Planck_2018_cosmo} evaluated in the no-bubble limit of our model, \(\beta=1\). This choice corresponds to a homogeneous \(\Lambda\)CDM cosmology and provides a consistent reference against which to compare the bubble prediction. Since \(\alpha_{\parallel}\) depends on the product \(H(z)r_d\), changes in the inferred line-of-sight expansion rate are partially degenerate with changes in the sound horizon scale. We therefore also show the ratio
\begin{equation}
    \frac{r_d^{\rm model}}{r_d^{\rm fid}}
\end{equation}
in Fig.~\ref{fig:rd_heatmap} to isolate the contribution from the varying BAO ruler.

To match the way the DESI BAO measurements are reported, we average the model prediction over redshift bins. We use a simple top-hat average,
\begin{equation}
    \bar{\alpha}_\parallel(z_{\rm mid})
    =
    \frac{1}{\Delta z}
    \int_{z_{\rm min}}^{z_{\rm max}}
    \alpha_\parallel(z)\,dz,
    \quad z_{\rm mid} = \frac{z_{\rm min}+z_{\rm max}}{2}, \quad
    \Delta z = z_{\rm max}-z_{\rm min}.
    \label{eq:alpha_par_binned}
\end{equation}
In particular, we use bins of width $\Delta z = 0.2$ for the two lowest-redshift bins, where the sharpest bubble-induced features appear. This binning is important because a localized feature in \(\alpha_\parallel(z_{\rm obs})\) would not be observed at a single redshift in the DESI analysis; instead, it would be averaged over all galaxies contributing to the corresponding tomographic bin. The binned model prediction \(\bar{\alpha}_\parallel\) is therefore the appropriate quantity to compare with the DESI DR2 measurements~\cite{DESI_DR2}. For various bubble parameters, we plot the parallel stretch, see Fig.~\ref{fig:par_stretch}.

\subsubsection{Perpendicular BAO Stretch}
\label{sec:perp_bao}
\begin{figure}
    \centering
    \includegraphics[width=\linewidth]{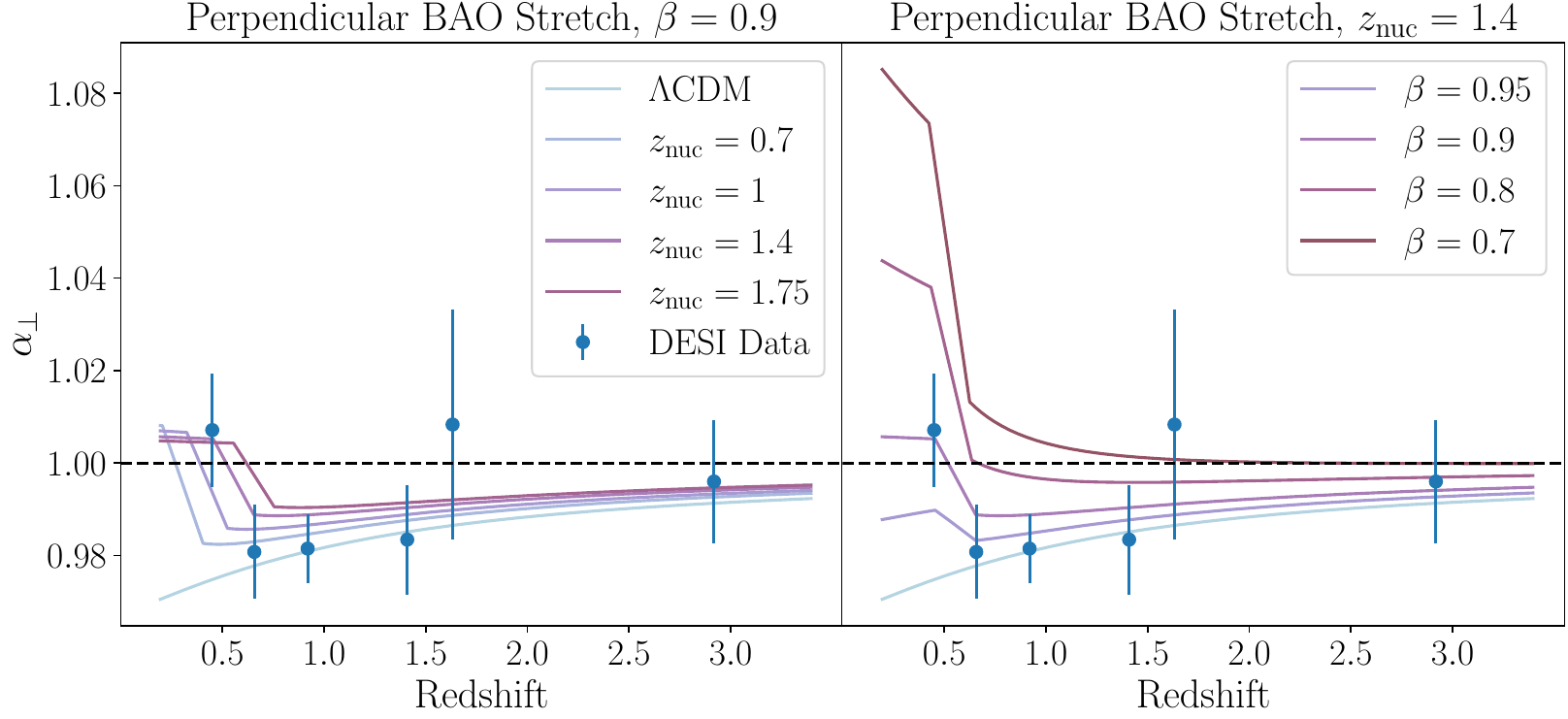}
    \caption{The Alcock--Paczynski perpendicular stretch parameters are plotted with varying $\beta$ and $z_{\rm nuc}$ values.}
    \label{fig:per_stretch}
\end{figure}
We also compute the perpendicular BAO stretch parameter, \(\alpha_\perp\), in the bubble model and compare it with the DESI BAO measurements. The perpendicular stretch quantifies the change in the inferred BAO scale transverse to the line of sight relative to the fiducial cosmology used in the BAO analysis. Again, following the Alcock--Paczynski stretch convention~\cite{Alam_2017,Beutler_2011,Alam_2021,DESI_DR1, DESI_DR2}, we define
\begin{equation}
    \alpha_\perp(z_{\rm obs})
    =
    \frac{
    D_M^{\rm model}(z_{\rm obs})\, r_d^{\rm fid}
    }{
    D_M^{\rm fid}(z_{\rm obs})\, r_d^{\rm model}
    }.
    \label{eq:alpha_perp}
\end{equation}
In Eq.~\eqref{eq:alpha_perp}, \(D_M^{\rm model}(z_{\rm obs})\) is the transverse comoving distance inferred from the bubble-model distance--redshift relation, as described in Sec.~\ref{sec:redshift}. For a central observer in the bubble, this is the comoving distance associated with the angular-diameter distance relation discussed in Sec.~\ref{sec:co-move_proper_ang_diam}. The quantity $r_d^{\rm model}$ is the drag-epoch sound horizon expressed in the appropriate comoving coordinates for the observed shell, as described in Sec.~\ref{sec:sound_horizon}.

As in Sec.~\ref{sec:hubble}, the fiducial quantities are evaluated in the no-bubble limit of the model, \(\beta=1\), using the Planck 2018 cosmological parameters~\cite{Planck_2018_cosmo}. This gives a homogeneous \(\Lambda\)CDM reference model against which the bubble prediction can be compared.

As in the previous section, Sec.~\ref{sec:par_bao}, we use the same top-hat average with bins of width $\Delta z = 0.2$ for the two lowest-redshift bins,
\begin{equation}
    \bar{\alpha}_\perp(z_{\rm mid})
    =
    \frac{1}{\Delta z}
    \int_{z_{\rm min}}^{z_{\rm max}}
    \alpha_\perp(z)\,dz,
    \quad z_{\rm mid} = \frac{z_{\rm min}+z_{\rm max}}{2}, \quad
    \Delta z = z_{\rm max}-z_{\rm min}.
    \label{eq:alpha_perp_binned}
\end{equation}
We plot the perpendicular stretch for various bubble parameters in Fig.~\ref{fig:per_stretch}.

\subsection{Additional CMB Constraints}
Additional CMB constraints, including the kSZ auto-power spectrum, the bubble-induced remote dipole field, and the sound horizon size, can be used to constrain the two remaining free parameters describing the bubble, $\beta$ and $z_{\rm nuc}$.
\label{sec:additional_cmb}
\begin{figure}
    \centering
    \includegraphics[width=0.9\linewidth]{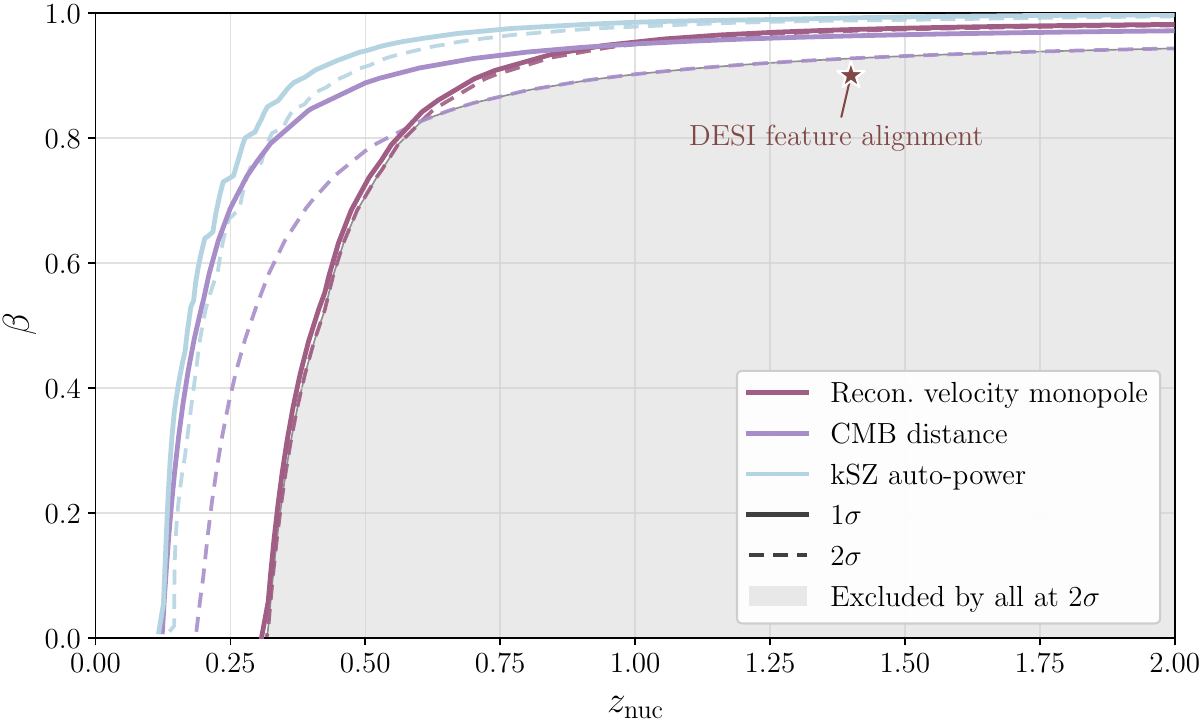}
    \caption{Constraints on $\beta$ and $z_{\rm nuc}$ derived from the additional CMB observations whose errors are assumed to be Gaussian. The shaded region is ruled out to 2$\sigma$ by all three constraints. The star represents the location of the parameter space at which the bubble model agrees qualitatively with the DESI data.}
    \label{fig:CMB_combined_constraint}
\end{figure}

\subsubsection{kSZ Constraints}
\label{sec:ksz_constraints}

The kSZ provides an additional constraint on our bubble parameters through the remote CMB dipole, the CMB dipole as seen by remote observers projected onto our line of sight. Although the observed CMB dipole constrains the Milky Way to lie close to the bubble centre, other points within our past light cone are not near the centre. Free electrons at different locations inside the bubble therefore observe different CMB dipoles because photons arriving from different directions cross the bubble wall at different times. This is illustrated in Fig. \ref{fig:ksz}. Thomson scattering of this locally anisotropic radiation field generates a kSZ contribution to the observed CMB anisotropy~\cite{original_SZ,Birkinshaw_1999}. 

Ref.~\cite{Pen_2014} first computed the kSZ signal for a one-bubble spacetime, demonstrating that the observed temperature power spectrum imposes a strong constraint. Here, we make similar predictions in the context of our model. We compute the bubble-induced remote dipole by repeating the CMB dipole calculation of Sec.~\ref{sec:cmb_dipole_constraints} for observers located at different positions and times along our past light cone. This gives an effective apparent velocity, $v_{\rm dip}(r)$, as computed in Sec.~\ref{sec:cmb_dipole_constraints} where \(r\) denotes the comoving distance to the scattering location along the observer's past light cone. This velocity should not be interpreted as a true peculiar velocity of matter. Rather, it is the effective velocity that would reproduce the same dipolar CMB anisotropy seen by the remote scatterer. The contribution is therefore sourced by the bubble-induced redshift anisotropy, not by ordinary peculiar motion or by the local Hubble flow.

To estimate the corresponding contribution to the CMB auto-power spectrum, we follow Ref.~\cite{matt_ksz2024} and write
\begin{multline}
    C_\ell^{vv}
    =
    \int dr\,\bigg(
    \sigma_T^2 \bar{n}_{e,0}^2
    \left(1+z(r)\right)^4
    \left(\frac{1}{H(z(r))}\right)^2\times\\
    b_e^2\!\left(z(r), r, \ell\right)
    v_{\rm dip}^2(r)
    P_k\!\left(z(r), \frac{\ell+1/2}{r}\right)\bigg).
    \label{eq:ksz_clvv}
\end{multline}
Here \(\sigma_T\) is the Thomson scattering cross section, \(\bar{n}_{e,0}\) is the mean electron number density today, \(b_e\) is the scale-dependent electron bias, all defined in  Ref.~\cite{matt_ksz2024}. We use the same parameter choices as Ref.~\cite{matt_ksz2024} and find $\sigma_T \bar{n}_{e,0} \approx 4.08\times10^{-7}{\rm Mpc}^{-1}$. \(P_k\) is the matter power spectrum, which we compute using CAMB with the DESI exterior cosmology density parameters mentioned in Sec.~\ref{sec:cosmo_param}. Since the bubble-induced remote dipole is present only inside the bubble and vanishes outside it, we do not distinguish between \(r_-\) and \(r_+\) in Eq.~\eqref{eq:ksz_clvv}.

\begin{figure}
    \centering
    \includegraphics[width=0.9\linewidth]{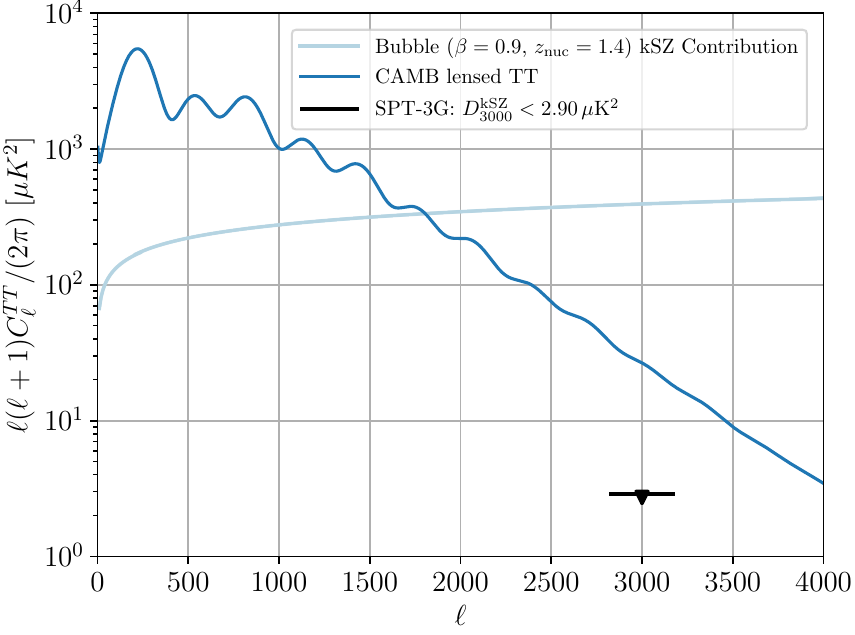}
    \caption{Bubble induced kinetic Sunyaev--Zel'dovich (kSZ) effect contribution to the TT autocorrelation CMB power spectrum for $\beta=0.9$ and $z_{\rm nuc}=1.4$. The 1$\sigma$ upper bound of the kSZ amplitude stated in Ref.~\cite{SPT} is shown in black. For these specific bubble parameters, the bubble induced contribution to the kSZ amplitude exceeds the expected upper bound to the kSZ contribution to the CMB power spectrum by two orders of magnitude, and the CMB power spectrum itself (plotted line created with CAMB) at $\ell=3000$ by one order of magnitude. This plot illustrates that the bubble of interest is strongly ruled out due to having a kSZ contribution that would dominate the CMB power spectrum in a way that is not observed.}
    \label{fig:kSZ_d_ell}
\end{figure}

Fig.~\ref{fig:kSZ_d_ell} shows the resulting bubble-induced kSZ contribution. Over the range of multipoles relevant for current measurements, the contribution increases monotonically with \(\ell\). Using constraints reported in Table X of Ref.~\cite{SPT} that the amplitude of the kSZ contribution to the CMB should be less than 2.90 (at 1$\sigma$) at an $\ell$ of 3000, we place bounds on $z_{\rm nuc}$ and $\beta$. The bounds can be seen in the Fig. \ref{fig:CMB_combined_constraint}. Notably, the kSZ effect is the most constraining, nearly ruling out the bubble entirely at high enough values of $z_{\rm nuc}$.

\subsubsection{Velocity Monopole Constraints}
\label{sec:velocity_monopole_constraints}

To compare with the velocity-monopole measurement of Ref.~\cite{matt_ksz2024}, we apply the same radial weighting used in their Planck $\times$ unWISE-blue reconstruction. The relevant observable is therefore not the apparent velocity at a single scattering location, but the window-averaged remote dipole,
\begin{equation}
    \langle v\rangle
    =
    \int dr\, W_v(r)\, v(r),
    \label{eq:velocity_monopole}
\end{equation}
where $W_v(r)$ is the velocity window function defined in Ref.~\cite{matt_ksz2024}, covering a redshift range $0.2 \lesssim z \lesssim 1$.

We compare this prediction to the monopole constraints reported in Table~II of Ref.~\cite{matt_ksz2024}. We use the 143 GHz Planck x unWISE velocity reconstruction as our fiducial constraint since this combination has the smallest measured monopole among the individual-frequency reconstructions and is less affected by the foreground monopole contamination that appears at other frequencies. For this reconstruction, Table~II of Ref.~\cite{matt_ksz2024} gives
\begin{equation}
M_{143}=3.81\times10^{-4},
\qquad
N_{143}=1.35\times10^{-8},
\end{equation}
in units of $v/c$ and $(v/c)^2$, respectively. We estimate the monopole uncertainty from the reconstruction noise as
\begin{equation}
\sigma_{\rm mono}
=
\sqrt{\frac{N_{143}}{4\pi}}.
\end{equation}
This gives an additional bound on $z_{\rm nuc}$ and $\beta$ as plotted in Fig.~\ref{fig:CMB_combined_constraint}.

\subsubsection{Sound Horizon Size Constraints}
\label{sec:sound_horizon_constraints}

We next use the sound horizon calibration discussed in Sec.~\ref{sec:sound_horizon} to constrain the bubble parameters; the results are shown in Fig.~\ref{fig:CMB_combined_constraint}. The key relation is
\begin{equation}
    r_s = \theta_s r_+^{\rm ls},
    \label{eq:rs_constraint_relation}
\end{equation}
where \(\theta_s\) is the angular size of the CMB acoustic scale, \(r_s\) is the comoving sound horizon at last scattering, and \(r_+^{\rm ls}\) is the exterior comoving distance to the last-scattering surface. In the bubble model, \(r_+^{\rm ls}\) is computed using the modified distance--redshift relation, with the observed last-scattering redshift fixed to \(z_{\rm ls}=1089.92\).

This relation provides a constraint because \(\theta_s\) is measured very precisely, while \(r_s\) is independently constrained by early-universe physics. For each choice of bubble parameters, we compute the distance \(r_+^{\rm ls}\), infer the corresponding value of \(r_s\) from Eq.~\eqref{eq:rs_constraint_relation}, and require that this inferred value remain consistent with the observationally allowed range of \(r_s\). Parameter choices that shift the bubble-model distance to last scattering too far therefore require an incompatible value of the sound horizon and are excluded.

Throughout this calculation, both \(r_s\) and \(r_+^{\rm ls}\) are treated as comoving quantities in the exterior coordinate system. This is appropriate because the last-scattering surface lies outside the bubble in the scenarios considered here, and because the physical acoustic scale is set before the late-time phase transition. As discussed in Sec.~\ref{sec:sound_horizon}, the conversion between comoving outer sound horizon size and the comoving inner sound horizon size has negligible impact on the resulting constraints.

\subsubsection{Consequence of Constraints}
\label{sec:consequence_of_constraints}
The preceding sections show that the bubble model has two competing requirements. To produce BAO features comparable to those seen by DESI, the bubble must be large enough and have a sufficiently strong dark-energy contrast to affect photons over the DESI redshift range. However, as we showed in this work, such bubbles also produce large anisotropies in the CMB through the dipole, remote-dipole, velocity-monopole, and sound horizon constraints. The region of parameter space that gives the most DESI-like features is therefore excluded by CMB data.

Fig.~\ref{fig:CMB_combined_constraint} shows the region of parameter space ruled out by the combined CMB constraints. The allowed models correspond primarily to late-nucleating bubbles, whose present-day sizes are small enough that CMB photons reaching us are not significantly affected by the wall. These models evade the CMB constraints, but at the cost of moving the bubble-induced features to redshifts too low to be useful in explaining the current DESI tension.

Thus, within the assumptions of this model, the CMB constraints prevent the bubble from producing an observable DESI-scale BAO signature. The broader interpretation of this result, and its implications for spatially varying dark-energy models more generally, are discussed in the following section.

\section{Discussion}

\label{sec:discussion}
The bubble model studied in this work provides a useful toy example of how spatial variation in dark energy can produce nontrivial signatures in cosmological distance measurements (as seen in Figs.~\ref{fig:r_vs_z}--\ref{fig:hubble},~\ref{fig:par_stretch}, and~\ref{fig:per_stretch}). In particular, the model naturally generates distinct features in the radial and transverse BAO observables, Figs.~\ref{fig:per_stretch} and~\ref{fig:par_stretch}, because photons may propagate through two FLRW regions and cross a moving boundary before reaching the observer. For some choices of the bubble nucleation redshift and dark-energy contrast, these features qualitatively resemble the Alcock--Paczynski distortions suggested by the DESI BAO measurements.

However, the same ingredients that make the model observationally interesting also make it strongly constrained. The dominant constraints arise from the CMB dipole, Fig.~\ref{fig:dipole_constraint}, and the kSZ effect, Fig.~\ref{fig:CMB_combined_constraint}. If the observer is located away from the centre of the bubble, the CMB redshift becomes direction dependent, producing an apparent dipole. The observed CMB dipole therefore forces the observer to lie close to the bubble centre. Even when this condition is imposed, remote observers along our past light cone and within the bubble generally see their own bubble-induced CMB dipoles, which generate a kSZ signal, illustrated in Fig.~\ref{fig:ksz}. Together with the velocity-monopole and sound horizon constraints, these effects rule out the region of parameter space that agrees best with the DESI BAO features.

Despite this negative result, the model is still informative. It demonstrates explicitly that spatially varying dark energy can generate features that are not easily mimicked by a purely homogeneous, time-varying equation of state. In particular, the bubble produces different behaviour in the parallel and perpendicular BAO stretch parameters, with sharp redshift dependence tied to the wall-crossing event. These qualitative signatures may be useful when thinking about broader classes of inhomogeneous or spacetime-dependent dark-energy models.

There are several important idealizations in the present construction. First, we have treated the bubble wall as a thin boundary and approximated its late-time trajectory as null. A more realistic cosmological phase transition may involve a finite wall thickness and a smoother interpolation in the dark energy density. Such effects could soften the sharp features in the distance--redshift relation and modify the associated CMB constraints.

Second, we have neglected possible interactions between the bubble wall and matter. In a more complete model, the passage of the wall through the matter distribution could perturb the local density field, generate peculiar velocities, alter the relation between matter and the dark-energy transition, or slow down the bubble wall to some subluminal terminal velocity. These effects would be important for understanding structure formation, galaxy clustering, and the detailed kSZ signal in a physically realistic scenario.

Third, we have assumed a perfectly spherical bubble. While this assumption makes the calculation tractable and isolates the basic observational effects, realistic bubbles may be deformed by interactions with matter, neighboring bubbles, or perturbations in the underlying field responsible for the transition. Departures from spherical symmetry could change the angular structure of the CMB and BAO signatures, and may lead to qualitatively different constraints.

Future work should therefore explore more general spacetime-dependent dark-energy models, including thick-wall transitions, subluminal wall propagation, matter-wall interactions, and nonspherical geometries. Although the simple bubble model studied in this paper is ruled out as an explanation of the DESI trend, it motivates a broader investigation of spatially varying dark energy as a possible source of distinctive cosmological distance signatures.

\section{Acknowledgments}
We thank Niayesh Afshordi for insight into this project, especially with regard to understanding the apparent velocity dipole. We also thank Jordan Krywonos for input and code related to the kSZ effect. We thank Cole Coughlin for discussions regarding the bubble wall Snell's Law effect. And we thank Elie Wolfe for a conversation regarding the paper's accessibility. MCJ and KJM are supported by the Natural Science and Engineering Research Council through a Discovery grant. We acknowledge the support of the Natural Sciences and Engineering Research Council of Canada (NSERC), RGPIN-03721-2023. This research was also supported in part by grant NSF PHY-2309135 to the Kavli Institute for Theoretical Physics (KITP) as well as the Perimeter Institute for Theoretical Physics. Research at Perimeter Institute is supported by the Government of Canada through the Department of Innovation, Science and Economic Development Canada and by the Province of Ontario through the Ministry of Colleges and Universities.

We acknowledge the use of AI in this paper for the purposes of literature searches, plotting, bug fixing, brainstorming discussions, edits to tone and clarity of writing, and extra help finding typos.

\bibliographystyle{JHEP_arxiv}
\bibliography{bibliography}

@article{Sakai_1994,
   title={Junction conditions of {Friedmann--Robertson--Walker} spacetimes},
   volume={50},
   doi={10.1103/physrevd.50.5425},
   number={8},
   journal={Phys. Rev. D},
   author={Sakai, Nobuyuki and Maeda, Kei-ichi},
   year={1994},
   pages={5425--5428} }

@article{Pen_2014,
   title={Observational consequences of dark energy decay},
   volume={89},
   doi={10.1103/physrevd.89.063009},
   number={6},
   journal={Phys. Rev. D},
   author={Pen, Ue-Li and Zhang, Pengjie},
   year={2014},
   pages={063009},}

@article{Pen_1998,
   title={Gamma-ray bursts from baryon decay in neutron stars},
   volume={509},
   doi={10.1086/306511},
   number={2},
   journal={Astrophys. J.},
   author={Pen, Ue-Li and Loeb, Abraham and Turok, Neil},
   year={1998},
   pages={537--543},
   eprint={astro-ph/9712178},
   archivePrefix={arXiv},
   primaryClass={astro-ph}}

@article{Coleman_1977,
  title = {Fate of the false vacuum: Semiclassical theory},
  author = {Coleman, Sidney},
  journal = {Phys. Rev. D},
  volume = {15},
  number = {10},
  pages = {2929--2936},
  year = {1977},
  doi = {10.1103/PhysRevD.15.2929},
}

@article{Linder_2003,
   title={Exploring the expansion history of the universe},
   volume={90},
   doi={10.1103/physrevlett.90.091301},
   number={9},
   journal={Phys. Rev. Lett.},
   author={Linder, Eric V.},
   year={2003},
   pages={091301},}

@article{Chevallier_2001,
   title={Accelerating universes with scaling dark matter},
   volume={10},
   doi={10.1142/s0218271801000822},
   number={02},
   journal={Int. J. Mod. Phys. D},
   author={Chevallier, Michel and Polarski, David},
   year={2001},
   pages={213--223} }

@article{Planck_2018_cosmo,
   title={{Planck} 2018 results. {VI}. Cosmological parameters},
   volume={641},
   doi={10.1051/0004-6361/201833910},
   journal={Astron. Astrophys.},
   author={Aghanim, N. and others},
   year={2020},
   pages={A6} }

@article{Planck_2018_dipole,
   title={{Planck} 2018 results. {I}. Overview and the cosmological legacy of {Planck}},
   volume={641},
   doi={10.1051/0004-6361/201833880},
   journal={Astron. Astrophys.},
   author={Aghanim, N. and others},
   year={2020},
   pages={A1} }

@article{Fixsen_2009,
   title={The temperature of the cosmic microwave background},
   volume={707},
   doi={10.1088/0004-637x/707/2/916},
   number={2},
   journal={Astrophys. J.},
   author={Fixsen, D. J.},
   year={2009},
   pages={916--920} }

@misc{DESI_DR1,
      title={Data Release 1 of the {Dark Energy Spectroscopic Instrument}}, 
      author={{DESI Collaboration} and Abdul-Karim, M. and others},
      year={2025},
      eprint={2503.14745},
      archivePrefix={arXiv},
      primaryClass={astro-ph.CO}, 
}

@article{DESI_DR2,
   title={{DESI} {DR2} results. {II.} Measurements of baryon acoustic oscillations and cosmological constraints},
   volume={112},
   doi={10.1103/tr6y-kpc6},
   number={8},
   journal={Phys. Rev. D},
   author={Abdul-Karim, M. and others},
   year={2025}}

@misc{Union3,
      title={Union through {UNITY}: Cosmology with 2,000 {SNe} using a unified Bayesian framework}, 
      author={Rubin, David and others},
      year={2025},
      eprint={2311.12098},
      archivePrefix={arXiv},
      primaryClass={astro-ph.CO}, 
}

@article{Pantheon+,
   title={The {Pantheon+} analysis: Cosmological constraints},
   volume={938},
   doi={10.3847/1538-4357/ac8e04},
   number={2},
   journal={Astrophys. J.},
   author={Brout, Dillon and others},
   year={2022},
   pages={110} }

@misc{DESY5,
      title={The {Dark Energy Survey}: Cosmology results with 1500 new high-redshift Type {Ia} supernovae using the full five-year dataset}, 
      author={{DES Collaboration} and Abbott, T. M. C. and others},
      year={2025},
      eprint={2401.02929},
      archivePrefix={arXiv},
      primaryClass={astro-ph.CO}, 
}

@misc{poisson2002,
      title={A reformulation of the {Barrab\`es--Israel} null-shell formalism}, 
      author={Poisson, Eric},
      year={2002},
      eprint={gr-qc/0207101},
      archivePrefix={arXiv},
      primaryClass={gr-qc}, 
}

@misc{matt_ksz2024,
      title={Kinetic {Sunyaev--Zel'dovich} velocity reconstruction from {Planck} and {unWISE}}, 
      author={Bloch, Richard and Johnson, Matthew C.},
      year={2024},
      eprint={2405.00809},
      archivePrefix={arXiv},
      primaryClass={astro-ph.CO}, 
}

@misc{act_dr6,
      title={The {Atacama Cosmology Telescope}: {DR6} power spectra, likelihoods, and {$\Lambda$CDM} parameters}, 
      author={Louis, Thibaut and others},
      year={2025},
      eprint={2503.14452},
      archivePrefix={arXiv},
      primaryClass={astro-ph.CO}, 
}

@article{SPT,
   title={{SPT-3G} {D1}: {CMB} temperature and polarization power spectra and cosmology from 2019 and 2020 observations of the {SPT-3G} main field},
   volume={113},
   doi={10.1103/7wt3-9v2y},
   number={8},
   journal={Phys. Rev. D},
   author={Camphuis, E. and others},
   year={2026}}

@article{Alam_2017,
   title={The clustering of galaxies in the completed {SDSS-III} Baryon Oscillation Spectroscopic Survey: Cosmological analysis of the {DR12} galaxy sample},
   volume={470},
   doi={10.1093/mnras/stx721},
   number={3},
   journal={Mon. Not. Roy. Astron. Soc.},
   author={Alam, Shadab and others},
   year={2017},
   pages={2617--2652} }

@article{Beutler_2011,
   title={The {6dF} Galaxy Survey: Baryon acoustic oscillations and the local {Hubble} constant},
   volume={416},
   doi={10.1111/j.1365-2966.2011.19250.x},
   number={4},
   journal={Mon. Not. Roy. Astron. Soc.},
   author={Beutler, Florian and others},
   year={2011},
   pages={3017--3032} }

@article{Alam_2021,
   title={Completed {SDSS-IV} extended Baryon Oscillation Spectroscopic Survey: Cosmological implications from two decades of spectroscopic surveys at the {Apache Point Observatory}},
   volume={103},
   doi={10.1103/physrevd.103.083533},
   number={8},
   journal={Phys. Rev. D},
   author={Alam, Shadab and others},
   year={2021},
   pages={083533},}

@article{original_SZ,
       author = {{Sunyaev}, R.~A. and {Zeldovich}, Ya. B.},
        title = "{The observations of relic radiation as a test of the nature of {X}-ray radiation from the Clusters of Galaxies}",
      journal = {Comments on Astrophysics and Space Physics},
         year = 1972,
       volume = {4},
        pages = {173},
}

@article{Birkinshaw_1999,
   title={The {{Sunyaev--Zel'dovich}} effect},
   volume={310},
   doi={10.1016/s0370-1573(98)00080-5},
   number={2-3},
   journal={Phys. Rept.},
   author={Birkinshaw, M.},
   year={1999},
   pages={97--195} }

@article{guth_could_1983,
    title = {Could the universe have recovered from a slow first-order phase transition?},
    volume = {212},
    doi = {10.1016/0550-3213(83)90307-3},
    number = {2},
    journal = {Nucl. Phys. B},
    author = {Guth, Alan H. and Weinberg, Erick J.},
    year = {1983},
    pages = {321--364},
}

@misc{aguirre_eternal_2007,
    title = {Eternal inflation, past and future},
    doi = {10.48550/arXiv.0712.0571},
    author = {Aguirre, Anthony},
    year = {2007},
    eprint = {0712.0571},
    archivePrefix = {arXiv},
    primaryClass = {hep-th},
}

@article{markkanen_cosmological_2018,
    title = {Cosmological aspects of {Higgs} vacuum metastability},
    volume = {5},
    doi = {10.3389/fspas.2018.00040},
    journal = {Front. Astron. Space Sci.},
    author = {Markkanen, Tommi and Rajantie, Arttu and Stopyra, Stephen},
    year = {2018},
    eprint = {1809.06923},
    archivePrefix = {arXiv},
    primaryClass = {astro-ph.CO},
    pages = {40},
}

@article{callan_fate_1977,
    title = {Fate of the false vacuum. {II}. First quantum corrections},
    volume = {16},
    doi = {10.1103/PhysRevD.16.1762},
    number = {6},
    journal = {Phys. Rev. D},
    author = {Callan, Curtis G. and Coleman, Sidney},
    year = {1977},
    pages = {1762--1768},
}

@article{coleman_gravitational_1980,
    title = {Gravitational effects on and of vacuum decay},
    volume = {21},
    doi = {10.1103/PhysRevD.21.3305},
    number = {12},
    journal = {Phys. Rev. D},
    author = {Coleman, Sidney and De Luccia, Frank},
    year = {1980},
    pages = {3305--3315},
}

@article{buttazzo_investigating_2013,
    title = {Investigating the near-criticality of the {Higgs} boson},
    volume = {2013},
    doi = {10.1007/JHEP12(2013)089},
    number = {12},
    journal = {JHEP},
    author = {Buttazzo, Dario and Degrassi, Giuseppe and Giardino, Pier Paolo and Giudice, Gian F. and Sala, Filippo and Salvio, Alberto and Strumia, Alessandro},
    year = {2013},
    eprint = {1307.3536},
    archivePrefix = {arXiv},
    primaryClass = {hep-ph},
    pages = {89},
}

@article{Bousso:2000xa,
    author = "Bousso, Raphael and Polchinski, Joseph",
    title = "{Quantization of four-form fluxes and dynamical neutralization of the cosmological constant}",
    eprint = "hep-th/0004134",
    archivePrefix = "arXiv",
    reportNumber = "SU-ITP-00-12, NSF-ITP-00-40",
    doi = "10.1088/1126-6708/2000/06/006",
    journal = "JHEP",
    volume = "06",
    pages = "006",
    year = "2000"
}

@misc{Susskind:2003kw,
    author = "Susskind, Leonard",
    editor = "Carr, Bernard J.",
    title = "{The anthropic landscape of string theory}",
    eprint = "hep-th/0302219",
    archivePrefix = "arXiv",
    pages = "247--266",
    year = "2003"
}

@article{RevModPhys.61.1,
  title = {The cosmological constant problem},
  author = {Weinberg, Steven},
  journal = {Rev. Mod. Phys.},
  volume = {61},
  number = {1},
  pages = {1--23},
  year = {1989},
  doi = {10.1103/RevModPhys.61.1},
}

@article{Giddings:2003zw,
    author = "Giddings, Steven B.",
    title = "{The fate of four dimensions}",
    eprint = "hep-th/0303031",
    archivePrefix = "arXiv",
    reportNumber = "MIFP-03-03",
    doi = "10.1103/PhysRevD.68.026006",
    journal = "Phys. Rev. D",
    volume = "68",
    pages = "026006",
    year = "2003"
}

@article{Giddings:2004vr,
    author = "Giddings, Steven B. and Myers, Robert C.",
    title = "{Spontaneous decompactification}",
    eprint = "hep-th/0404220",
    archivePrefix = "arXiv",
    doi = "10.1103/PhysRevD.70.046005",
    journal = "Phys. Rev. D",
    volume = "70",
    pages = "046005",
    year = "2004"
}

@article{Page:2006dt,
    author = "Page, Don N.",
    title = "{Is our universe likely to decay within 20 billion years?}",
    eprint = "hep-th/0610079",
    archivePrefix = "arXiv",
    reportNumber = "ALBERTA-THY-08-06",
    doi = "10.1103/PhysRevD.78.063535",
    journal = "Phys. Rev. D",
    volume = "78",
    pages = "063535",
    year = "2008"
}

@article{Brown:1988kg,
    author = "Brown, J. David and Teitelboim, C.",
    title = "{Neutralization of the cosmological constant by membrane creation}",
    doi = "10.1016/0550-3213(88)90559-7",
    journal = "Nucl. Phys. B",
    volume = "297",
    pages = "787--836",
    year = "1988"
}

@article{Abbott:1984qf,
    author = "Abbott, L. F.",
    title = "{A mechanism for reducing the value of the cosmological constant}",
    reportNumber = "BRX-TH-175",
    doi = "10.1016/0370-2693(85)90459-9",
    journal = "Phys. Lett. B",
    volume = "150",
    pages = "427--430",
    year = "1985"
}

@article{Feng_2001,
   title={Saltatory relaxation of the cosmological constant},
   volume={602},
   doi={10.1016/s0550-3213(01)00097-9},
   number={1-2},
   journal={Nucl. Phys. B},
   author={Feng, Jonathan L. and March-Russell, John and Sethi, Savdeep and Wilczek, Frank},
   year={2001},
   pages={307--328} }

@article{Degrassi:2012ry,
    author = "Degrassi, Giuseppe and Di Vita, Stefano and Elias-Miro, Joan and Espinosa, Jose R. and Giudice, Gian F. and Isidori, Gino and Strumia, Alessandro",
    title = "{{Higgs} mass and vacuum stability in the {Standard Model} at {NNLO}}",
    eprint = "1205.6497",
    archivePrefix = "arXiv",
    primaryClass = "hep-ph",
    reportNumber = "CERN-PH-TH-2012-134, RM3-TH-12-9",
    doi = "10.1007/JHEP08(2012)098",
    journal = "JHEP",
    volume = "08",
    pages = "098",
    year = "2012"
}

@article{Isidori:2001bm,
    author = "Isidori, Gino and Ridolfi, Giovanni and Strumia, Alessandro",
    title = "{On the metastability of the {Standard Model} vacuum}",
    eprint = "hep-ph/0104016",
    archivePrefix = "arXiv",
    reportNumber = "CERN-TH-2001-092, LNF-01-014-P, GEF-TH-6-01, IFUP-TH-2001-11",
    doi = "10.1016/S0550-3213(01)00302-9",
    journal = "Nucl. Phys. B",
    volume = "609",
    pages = "387--409",
    year = "2001"
}

@article{Cabibbo:1979ay,
    author = "Cabibbo, N. and Maiani, L. and Parisi, G. and Petronzio, R.",
    title = "{Bounds on the fermions and {Higgs}-boson masses in grand unified theories}",
    reportNumber = "CERN-TH-2683",
    doi = "10.1016/0550-3213(79)90167-6",
    journal = "Nucl. Phys. B",
    volume = "158",
    pages = "295--305",
    year = "1979"
}

@misc{Bai:2026sux,
    author = "Bai, Yang and Lu, Sida and Orlofsky, Nicholas",
    title = "{Late-time quantum vacuum decay and its cosmological implications}",
    eprint = "2605.30259",
    archivePrefix = "arXiv",
    primaryClass = "astro-ph.CO",
    year = "2026"
}

@article{Rudelius:2019cfh,
    author = "Rudelius, Tom",
    title = "{Conditions for (no) eternal inflation}",
    eprint = "1905.05198",
    archivePrefix = "arXiv",
    primaryClass = "hep-th",
    doi = "10.1088/1475-7516/2019/08/009",
    journal = "JCAP",
    volume = "08",
    pages = "009",
    year = "2019"
}

@misc{Banks:2005ru,
    author = "Banks, T. and Johnson, M.",
    title = "{Regulating eternal inflation}",
    eprint = "hep-th/0512141",
    archivePrefix = "arXiv",
    year = "2005"
}

@article{Aguirre:2006ap,
    author = "Aguirre, A. and Banks, T. and Johnson, M.",
    title = "{Regulating eternal inflation. {II}. The great divide}",
    eprint = "hep-th/0603107",
    archivePrefix = "arXiv",
    doi = "10.1088/1126-6708/2006/08/065",
    journal = "JHEP",
    volume = "08",
    pages = "065",
    year = "2006"
}

@article{Shafieloo:2016bpk,
    author = "Shafieloo, Arman and Hazra, Dhiraj Kumar and Sahni, Varun and Starobinsky, Alexei A.",
    title = "{Metastable dark energy with radioactive-like decay}",
    eprint = "1610.05192",
    archivePrefix = "arXiv",
    primaryClass = "astro-ph.CO",
    doi = "10.1093/mnras/stx2481",
    journal = "Mon. Not. Roy. Astron. Soc.",
    volume = "473",
    number = "2",
    pages = "2760--2770",
    year = "2018"
}

@article{Li:2019san,
    author = "Li, Xiaolei and Shafieloo, Arman and Sahni, Varun and Starobinsky, Alexei A.",
    title = "{Revisiting metastable dark energy and tensions in the estimation of cosmological parameters}",
    eprint = "1904.03790",
    archivePrefix = "arXiv",
    primaryClass = "astro-ph.CO",
    doi = "10.3847/1538-4357/ab535d",
    journal = "Astrophys. J.",
    volume = "887",
    pages = "153",
    year = "2019"
}

@ARTICLE{Blau_1987,
  title = {Dynamics of false-vacuum bubbles},
  author = {Blau, Steven K. and Guendelman, E. I. and Guth, Alan H.},
  journal = {Phys. Rev. D},
  volume = {35},
  number = {6},
  pages = {1747--1766},
  year = {1987},
  doi = {10.1103/PhysRevD.35.1747},
}

@article{Aurilia:1989sb,
    author = "Aurilia, Antonio and Palmer, Marc and Spallucci, Euro",
    title = "{Evolution of bubbles in a vacuum}",
    reportNumber = "SLAC-PUB-4697, SLAC-PUB-4697-REV",
    doi = "10.1103/PhysRevD.40.2511",
    journal = "Phys. Rev. D",
    volume = "40",
    pages = "2511",
    year = "1989"
    }

@article{Sato_1986,
    author = {Sato, Humitaka},
    title = {Motion of a shell at metric junction},
    journal = {Prog. Theor. Phys.},
    volume = {76},
    number = {6},
    pages = {1250--1259},
    year = {1986},
    doi = {10.1143/PTP.76.1250},
}

@article{Berezin_1987,
  title = {Dynamics of bubbles in general relativity},
  author = {Berezin, V. A. and Kuzmin, V. A. and Tkachev, I. I.},
  journal = {Phys. Rev. D},
  volume = {36},
  number = {10},
  pages = {2919--2944},
  year = {1987},
  doi = {10.1103/PhysRevD.36.2919},
}

@article{Aguirre_Matt_2005,
   title={Dynamics and instability of false vacuum bubbles},
   volume={72},
   doi={10.1103/physrevd.72.103525},
   number={10},
   journal={Phys. Rev. D},
   author={Aguirre, Anthony and Johnson, Matthew C.},
   year={2005},
   pages={103525},}

@article{Aguirre_Matt_2006,
   title={Two tunnels to inflation},
   volume={73},
   doi={10.1103/physrevd.73.123529},
   number={12},
   journal={Phys. Rev. D},
   author={Aguirre, Anthony and Johnson, Matthew C.},
   year={2006},
   pages={123529},}

@article{Johnson_2020,
   title={Stable thin-wall negative-mass bubbles in de {Sitter} spacetime},
   volume={52},
   doi={10.1007/s10714-020-02732-9},
   number={8},
   journal={Gen. Rel. Grav.},
   author={Johnson, Matthew C. and Paranjape, M. B. and Savard, Antoine and Tapia-Arellano, Natalia},
   year={2020}}

@article{Marra_2013,
   title={Cosmic variance and the measurement of the local {Hubble} parameter},
   volume={110},
   doi={10.1103/physrevlett.110.241305},
   number={24},
   journal={Phys. Rev. Lett.},
   author={Marra, Valerio and Amendola, Luca and Sawicki, Ignacy and Valkenburg, Wessel},
   year={2013},
   pages={241305},}

@article{Garcia_Bellido_2008,
   title={Confronting {Lema{\^i}tre--Tolman--Bondi} models with observational cosmology},
   volume={2008},
   doi={10.1088/1475-7516/2008/04/003},
   number={04},
   journal={JCAP},
   author={Garc{\'i}a-Bellido, Juan and Haugb{\o}lle, Troels},
   year={2008},
   pages={003} }

@article{Alnes_2006,
   title={Inhomogeneous alternative to dark energy?},
   volume={73},
   doi={10.1103/physrevd.73.083519},
   number={8},
   journal={Phys. Rev. D},
   author={Alnes, H{\aa}vard and Amarzguioui, Morad and Gr{\o}n, {\O}yvind},
   year={2006},
   pages={083519},}

@article{Celerier2000,
  author        = {C{\'e}l{\'e}rier, Marie-No{\"e}lle},
  title         = {Do we really see a cosmological constant in the supernovae data?},
  journal       = {Astronomy \& Astrophysics},
  volume        = {353},
  pages         = {63--71},
  year          = {2000},
  eprint        = {astro-ph/9907206},
  archivePrefix = {arXiv},
  primaryClass  = {astro-ph},
}

\appendix

\section*{Appendix}

\section{Bubble Wall Velocity Derivation}
\label{append:bubble_wall_eom}

Using Eq.~(8) from Ref.~\cite{Sakai_1994} for a flat FLRW cosmology where $r = \chi$, we see that 
\begin{equation}
    \frac{dR}{d\tau}= \gamma \qty(v + HR)\Big|_\pm,
\end{equation}
where $\gamma = \frac{dt}{d\tau} = \frac{1}{\sqrt{1-v^2}}$. Note that $v = a(dr/dt)$ is the bubble wall's peculiar velocity relative to the background expansion, and we assume it is positive with respect to our coordinate system (it is moving radially outwards). We can use the relationship between $v$ and $\gamma$ to obtain $\frac{dR}{d\tau}$ fully in terms of $\gamma$, $H$, and $R$.

Taking the positive square root in the quadratic equation, we find
\begin{equation}
    \gamma = \frac{-H\dot{R}R +\sqrt{\dot{R}^2+1-H^2R^2}}{1-H^2R^2}
\end{equation}
where $\dot{R} = \frac{dR}{d\tau}$. Now to relate $\frac{dR}{dt}$ to $R_0$ and remove the dependence on $\frac{dR}{d\tau}$, we will need two more equations from Ref.~\cite{Sakai_1994}: Eq. (9) 
\begin{equation}
    \qty(K_\theta^{\theta\pm})^2 = \frac{1}{R^2}\qty{1+\qty(\frac{dR}{d\tau})^2-\frac{8\pi G \rho^\pm}{3}R^2}
\end{equation}
and Eq. (10)
\begin{equation}
    K_\theta^{\theta\pm} = \frac{\rho^+ - \rho^- \mp 6\pi G \sigma^2}{3\sigma}
\end{equation}
where $K_\theta^{\theta\pm}$ is the extrinsic curvature of the two spacetimes at the boundary, $\sigma$ is the energy density or tension of the bubble wall, and $\frac{8\pi G \rho^\pm}{3} = H^2_\pm $. If we make the assumption that at nucleation and within a short time period after nucleation, the bubble is deeply sub-horizon, i.e., $H_\pm R \ll 1$, and that for a short time period after nucleation, $\rho^\pm \approx$ const., then we can show that the bubble becomes relativistic on a negligible timescale.
To continue, we note that $\frac{dR}{d\tau}$ is zero at nucleation, and in the regime of a just nucleated, deeply sub-horizon bubble, $K_\theta^{\theta\pm}$ is constant. Additionally, in this regime, $H_\pm R_0^2 \ll 1$. Therefore we obtain
\begin{equation}
    \qty(K_{\theta}^{\theta\pm}) \approx \frac{1}{R_0^2} \approx \frac{1}{R^2}\qty{1+\qty(\frac{dR}{d\tau})^2-H^2_\pm R^2}
\end{equation}
and can solve for $\frac{dR}{d\tau}$ as a function of $R$, $R_0$, and $H_\pm$.

Next, combining $\gamma$ with $\frac{dR}{d\tau}$ lets us obtain $\frac{dR}{dt}$,
\begin{equation}
    \frac{dR}{dt} = \frac{dR}{d\tau}\qty(\frac{dt}{d\tau})^{-1} = \frac{dR}{d\tau}\gamma^{-1}.
\end{equation}
\noindent We find
\begin{equation}
    \frac{dR}{dt} = \sqrt{\frac{R^2}{R_0^2}+H^2_\pm R^2 -1}\qty(\frac{1-H_\pm^2R^2}{\frac{R}{R_0}-H_\pm R\sqrt{\qty(\frac{1}{R_0^2}+H_\pm^2)R^2-1}}).
\end{equation}
In the regime where $R$ is about one or two orders of magnitude larger than $R_0$ while $H_\pm R \ll 1$ or deeply subhorizon, we find that
\begin{equation}
    \frac{dR}{dt}\rightarrow 1,
\end{equation}
the bubble wall becomes ultra-relativistic, approaching the speed of light. 

Note we assume the matter density difference inside and outside the bubble (which is zero at nucleation but grows with time) is negligible. The matter-density difference only produces a transient reduction in the pressure driving the wall; it does not act as a frictional or dissipative force. Since the dark energy density jump remains positive, the wall continues to be driven outward even if the acceleration is slightly reduced at some times. Once the wall has already become ultra-relativistic, this reduced-but-still-positive driving is not capable of making it slow down; instead, the wall remains near light speed and continues becoming asymptotically closer to the speed of light. Thus the light-cone approximation should remain valid, provided the matter correction never reverses the effective pressure difference.

\section{Matter and Vacuum Energy Content}
\label{append:friedmann}
The exterior cosmology is specified at time \(t_+^0\) by \(H_{0,+}\), \(\Omega_{M,0}^{+}\), and
\(\Omega_{\Lambda,0}^{+}\), such that
\begin{align}
    \rho_{M,0}^{+}
    &=
    \Omega_{M,0}^{+}
    \frac{3\qty(H_{0,+})^2}{8\pi G},
    \\
    \rho_{\Lambda}^{+}
    &=
    \Omega_{\Lambda,0}^{+}
    \frac{3\qty(H_{0,+})^2}{8\pi G}.
\end{align}
\noindent We assume that the matter density is continuous at nucleation,
\begin{equation}
    \rho_{M}^-\qty(t_-^{\rm nuc})
    =
    \rho_{M}^+\qty(t_+^{\rm nuc})
    \equiv
    \rho_{M, \rm nuc},
    \label{eq:matter_density_nucleation}
\end{equation}
whereas the vacuum energy density decreases inside the bubble as
specified previously, such that
\(\rho_{\Lambda}^{-}=\beta\rho_{\Lambda}^{+}\).
The normalization of an FLRW scale factor is arbitrary. We choose the
interior and exterior scale factors to agree at nucleation:
\begin{equation}
    a_-\qty(t_-^{\rm nuc})
    =
    a_+\qty(t_+^{\rm nuc})
    \equiv
    a_{\rm nuc}
    =
    \frac{1}{1+z_{\rm nuc}},
    \label{eq:scale_factor_nucleation}
\end{equation}
where \(z_{\rm nuc}\) is defined using the homogeneous exterior
cosmology. Conservation of pressureless matter then gives
\begin{equation}
    \rho_{M}^{\pm}(t_\pm)
    =
    \rho_{M,\rm nuc}
    \qty(\frac{a_{\rm nuc}}{a_\pm(t_\pm)})^3.
\end{equation}
Since
\begin{equation}
    \rho_{M,\rm nuc}
    =
    \rho_{M,0}^{+}(1+z_{\rm nuc})^3
    =
    \rho_{M,0}^{+}a_{\rm nuc}^{-3},
\end{equation}
the matter density on either side may be written as
\begin{equation}
    \rho_{M}^\pm(t_\pm)
    =
    \rho_{M,0}^{+}a_\pm^{-3}(t_\pm).
    \label{eq:matter_density_both_regions}
\end{equation}
\noindent The interior Friedmann equation is therefore
\begin{align}
    H^2_{-}(t_-)
    &=
    \frac{8\pi G}{3}
    \qty(
        \rho_{M,0}^{+}a_-^{-3}
        +
        \beta\rho_{\Lambda}^{+}
    )
    \nonumber\\
    &=
    \qty(H_{0,+})^2
    \qty[
        \Omega_{M,0}^{+}a_-^{-3}
        +
        \beta\Omega_{\Lambda,0}^{+}
    ].
    \label{eq:interior_friedmann_appendix}
\end{align}
At nucleation, this becomes
\begin{equation}
    \frac{H_{\rm nuc,-}}{H_{0,+}}
    =
    \sqrt{
        \Omega_{M,0}^{+}(1+z_{\rm nuc})^3
        +
        \beta\Omega_{\Lambda,0}^{+}
    }.
    \label{eq:interior_hubble_nucleation}
\end{equation}
\noindent The corresponding interior density fractions at nucleation are
\begin{align}
    \Omega_{M,\rm nuc}^{-}
    &=
    \frac{
        \Omega_{M,0}^{+}(1+z_{\rm nuc})^3
    }{
        \Omega_{M,0}^{+}(1+z_{\rm nuc})^3
        +
        \beta\Omega_{\Lambda,0}^{+}
    },
    \\
    \Omega_{\Lambda,\rm nuc}^{-}
    &=
    \frac{
        \beta\Omega_{\Lambda,0}^{+}
    }{
        \Omega_{M,0}^{+}(1+z_{\rm nuc})^3
        +
        \beta\Omega_{\Lambda,0}^{+}
    }.
\end{align}
\noindent Choosing \(t_-=0\) at the extrapolated Big Bang of the interior FLRW
solution, Eq.~\eqref{eq:interior_friedmann_appendix} integrates to
\begin{equation}
    a_-(t_-)
    =
    \qty(
        \frac{
            \Omega_{M,0}^{+}
        }{
            \beta\Omega_{\Lambda,0}^{+}
        }
    )^{1/3}
    \sinh^{2/3}
    \qty[
        \frac{3}{2}
        H_{0,+}
        \sqrt{\beta\Omega_{\Lambda,0}^{+}}\,
        t_-
    ].
    \label{eq:scale_factor_in_in}
\end{equation}
For comparison, the exterior scale factor is obtained by setting
\(\beta=1\):
\begin{equation}
    a_+(t_+)
    =
    \qty(
        \frac{
            \Omega_{M,0}^{+}
        }{
            \Omega_{\Lambda,0}^{+}
        }
    )^{1/3}
    \sinh^{2/3}
    \qty[
        \frac{3}{2}
        H_{0,+}
        \sqrt{\Omega_{\Lambda,0}^{+}}\,
        t_+
    ].
\end{equation}

Although the functional form of \(a_-(t_-)\) does not depend explicitly
on \(z_{\rm nuc}\), the nucleation redshift determines the interior time
at which the bubble forms as can be seen by combining Eqs.~\eqref{eq:scale_factor_nucleation}
and~\eqref{eq:scale_factor_in_in} which gives
\begin{equation}
    t_-^{\rm nuc}
    =
    \frac{
        2
    }{
        3H_{0,+}
        \sqrt{\beta\Omega_{\Lambda,0}^{+}}
    }
    \operatorname{arcsinh}
    \qty[
        \sqrt{
            \frac{
                \beta\Omega_{\Lambda,0}^{+}
            }{
                \Omega_{M,0}^{+}
            }
        }
        (1+z_{\rm nuc})^{-3/2}
    ].
\end{equation}

\section{Angular-Diameter Distance Across the Null Wall}
\label{append:angular_diameter_dist_calc}

\begin{figure}
    \centering
    \includegraphics[width=0.9\linewidth]{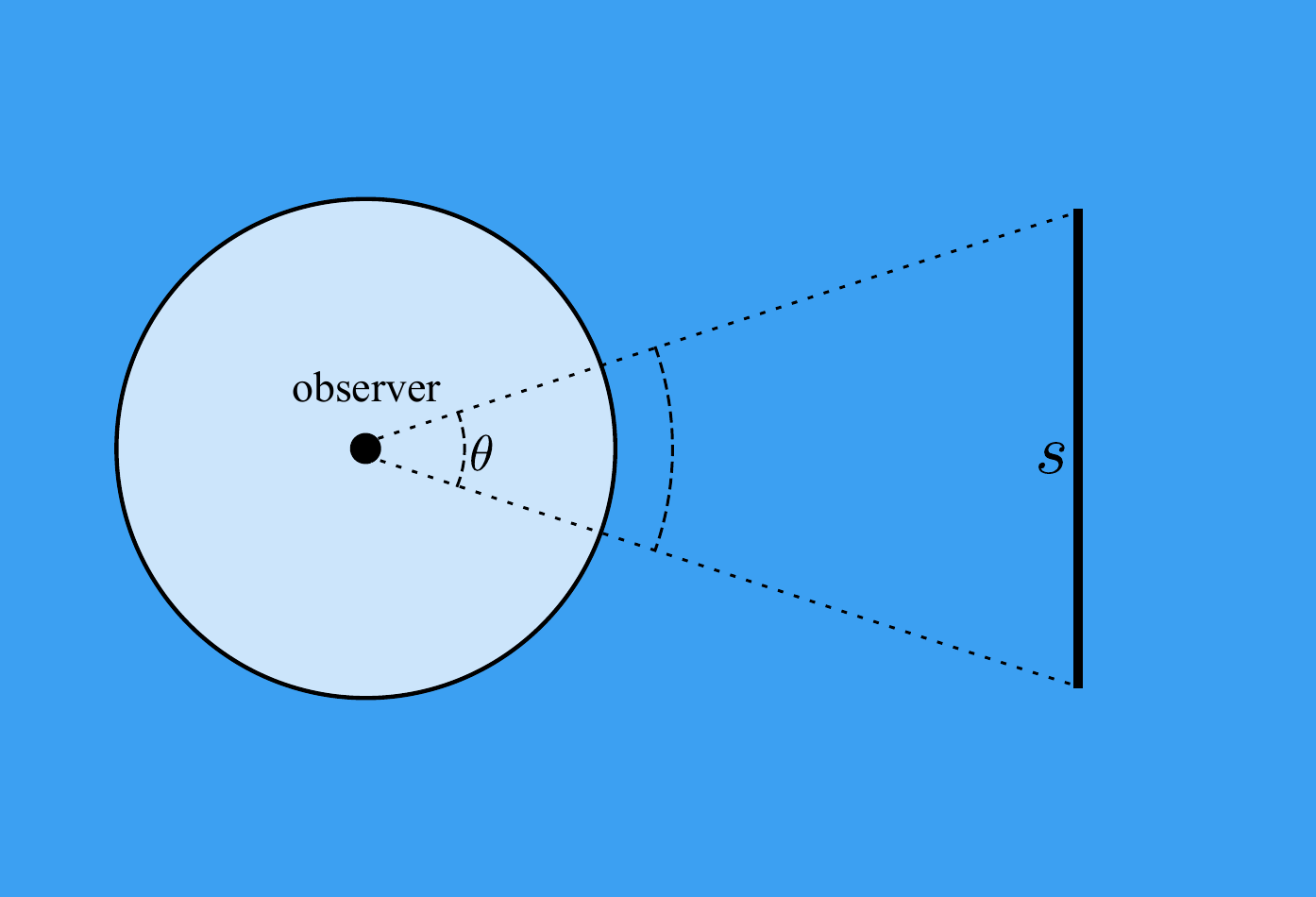}
    \caption{Illustration in comoving space of the geodesic deviation equation setup with geodesics originating from top and bottom of line segment, $s$, and converging at the observer at the bubble centre. The angle between the geodesics is derived to be the same angle inside the bubble as well as outside the bubble. Therefore the usual expression for angular-diameter distance equation from a flat FLRW cosmology holds.}
    \label{fig:angular_diameter_distance}
\end{figure}

We derive the angular-diameter distance formula for the setup where the observer is at the origin of the spacetime, located at the centre of the bubble, and light rays coming from the top and bottom of an object outside the bubble, length $s$, converge at the observer; see Fig.~\ref{fig:angular_diameter_distance}. Consider the propagation of an infinitesimal bundle of radial null geodesics across the bubble wall, and let \(\lambda_\gamma\) be an affine parameter along the photon geodesics. We denote the photon tangent by
\begin{equation}
    P^\alpha \equiv \frac{dx^\alpha}{d\lambda_\gamma}.
\end{equation}
This affine parameter, $\lambda_\gamma$, should be distinguished from the intrinsic parameter used to describe the null generators of the wall.

For a shear-free radial bundle, the geodesic-deviation equation may be written as
\begin{equation}
    \frac{d^2 s}{d\lambda_\gamma^2}
    =
    -\frac{1}{2}
    R_{\alpha\beta}P^\alpha P^\beta\,s .
    \label{eq:geod_dev_sachs}
\end{equation}
Using Einstein's equation and \(P^\alpha P_\alpha=0\), this becomes
\begin{equation}
    \frac{d^2 s}{d\lambda_\gamma^2}
    =
    -4\pi T_{\alpha\beta}P^\alpha P^\beta\,s .
    \label{eq:geod_dev_T}
\end{equation}

The only contribution which can produce a discontinuity in \(ds/d\lambda_\gamma\) is the distributional stress-energy localized on the null wall. Following the Barrabès--Israel/Poisson null-shell formalism \cite{poisson2002}, the wall contribution may be written as
\begin{equation}
    T^{\alpha\beta}_\Sigma
    =
    (-k_\mu u^\mu)^{-1}
    S^{\alpha\beta}\delta(\tau),
    \label{eq:poisson_shell_T}
\end{equation}
where \(k^\alpha\) is tangent to the null generators of the wall, \(u^\alpha\) is the four-velocity of an auxiliary timelike geodesic congruence intersecting the wall, and \(\tau\) is the proper time along that congruence. The parameter \(\tau\) is not a proper time of the wall itself, since the wall is null.

To express the distributional term along the photon trajectory, we use
\begin{equation}
    \delta(\tau)
    =
    \frac{\delta(\lambda_\gamma-\lambda_\gamma^0)}
    {d\tau/d\lambda_\gamma}\,,
\end{equation}
where \(\lambda_\gamma^0\) denotes the value of the photon affine parameter at the crossing. From Poisson's construction,
\begin{equation}
    \partial_\alpha \tau
    =
    \frac{k_\alpha}{k_\mu u^\mu},
\end{equation}
so that, along the photon path,
\begin{equation}
    \frac{d\tau}{d\lambda_\gamma}
    =
    P^\alpha \partial_\alpha \tau
    =
    \frac{k_\alpha P^\alpha}{k_\mu u^\mu}.
\end{equation}
Combining these expressions, the explicit dependence on the auxiliary congruence cancels, and the shell contribution to the focusing equation may be written as
\begin{equation}
    \frac{d^2s}{d\lambda_\gamma^2}
    =
    4\pi
    (k_\alpha P^\alpha)^{-1}
    S_{\alpha\beta}P^\alpha P^\beta
    \delta(\lambda_\gamma-\lambda_\gamma^0)\,s ,
    \label{eq:geod_dev_shell_general}
\end{equation}
up to the overall sign convention fixed by the orientation of \(k^\alpha\).

The surface stress tensor on the null wall can be decomposed \cite{poisson2002} as
\begin{equation}
    S^{\alpha\beta}
    =
    \mu k^\alpha k^\beta
    +
    j^A
    \left(
        k^\alpha e_A^\beta
        +
        e_A^\alpha k^\beta
    \right)
    +
    p\,\sigma^{AB}e_A^\alpha e_B^\beta ,
    \label{eq:S_decomp}
\end{equation}
where \(\mu\) is the surface density, \(j^A\) is the surface current, \(p\) is the surface pressure, and \(e_A^\alpha\) span the angular directions on the wall. For a radial photon bundle, \(P^\alpha e^A_\alpha=0\). Therefore the current and pressure terms do not contribute to \(S_{\alpha\beta}P^\alpha P^\beta\), and
\begin{equation}
    S_{\alpha\beta}P^\alpha P^\beta
    =
    \mu (k_\alpha P^\alpha)^2 .
\end{equation}
Equation \eqref{eq:geod_dev_shell_general} then reduces to
\begin{equation}
    \frac{d^2s}{d\lambda_\gamma^2}
    =
    4\pi \mu
    (k_\alpha P^\alpha)
    \delta(\lambda_\gamma-\lambda_\gamma^0)\,s .
    \label{eq:geod_dev_shell_mu}
\end{equation}
\noindent We choose the intrinsic parameter along the wall to be the areal radius
\begin{equation}
    \lambda = R,
    \qquad
    R \equiv a_\pm(t_\pm)r_\pm .
\end{equation}
This is a convenient choice because \(R\) is continuous across the junction, even though the comoving coordinates \(r_+\) and \(r_-\) are not. With this choice,
\begin{equation}
    k_\pm^\alpha
    =
    \frac{dx_\pm^\alpha}{dR}
    =
    \left(
        \frac{dt_\pm}{dR},
        \frac{dr_\pm}{dR},
        0,
        0
    \right).
\end{equation}
For an outgoing null wall in a spatially flat FLRW region,
\begin{equation}
    \frac{dR}{dt_\pm}
    =
    H_\pm R + 1,
\end{equation}
and therefore
\begin{equation}
    k_\pm^\alpha
    =
    \left(
        \frac{1}{H_\pm R+1},
        \frac{1}{a_\pm(H_\pm R+1)},
        0,
        0
    \right).
    \label{eq:k_wall_R}
\end{equation}

\noindent The photon is taken to be radially ingoing. We write
\begin{equation}
    P_\pm^\alpha
    =
    \left(
        \omega_\pm,
        -\frac{\omega_\pm}{a_\pm},
        0,
        0
    \right),
    \qquad
    \omega_\pm
    \equiv
    \frac{dt_\pm}{d\lambda_\gamma}.
    \label{eq:P_photon}
\end{equation}
Using \eqref{eq:k_wall_R} and \eqref{eq:P_photon}, one obtains
\begin{equation}
    k^\pm_\alpha P^\alpha_\pm
    =
    -\frac{2\omega_\pm}{H_\pm R+1}.
    \label{eq:kdotP}
\end{equation}

\noindent As noted in Eq.~(2.9) of Ref.~\cite{poisson2002}, the scalar \(k_\alpha P^\alpha\) is continuous across the wall for the radial photon trajectory,
\begin{equation}
    [k_\alpha P^\alpha]
    \equiv
    k^-_\alpha P^\alpha_-
    -
    k^+_\alpha P^\alpha_+
    =
    0 .
    \label{eq:kP_continuity}
\end{equation}
Equivalently,
\begin{equation}
    \frac{\omega_-}{\omega_+}
    =
    \frac{H_-R+1}{H_+R+1}.
    \label{eq:omega_matching}
\end{equation}
This relation should not be interpreted as a redshift measured in the rest frame of the wall, since a null wall has no rest frame. Rather, it is a matching condition for the tangential covariant component of the photon momentum across the continuous null hypersurface.

For the same choice \(\lambda=R\), the surface density obtained from the jump in the transverse curvature is
\begin{equation}
    \mu
    =
    -\frac{1}{8\pi R}
    \left[
        (H_-R+1)(H_-R-1)
        -
        (H_+R+1)(H_+R-1)
    \right].
    \label{eq:mu_result}
\end{equation}
Substituting \eqref{eq:kdotP} and \eqref{eq:mu_result} into \eqref{eq:geod_dev_shell_mu}, and integrating across an infinitesimal interval containing the wall, gives the jump condition
\begin{equation}
    \frac{ds_-}{d\lambda_{\gamma}}\bigg{|}_- - \frac{ds_+}{d\lambda_{\gamma}}\bigg{|}_+=
    \frac{s(\lambda_\gamma^0)}{R}
    \left[
        \omega_-(H_-R-1)
        -
        \omega_+(H_+R-1)
    \right]_{\Sigma},
    \label{eq:s_jump}
\end{equation}
where all quantities on the right-hand side are evaluated at the crossing event and $\Sigma$ denotes the wall.

We now show that this jump condition is precisely consistent with the usual areal-radius expression for the separation of central radial rays. Along an ingoing photon in either FLRW region,
\begin{equation}
    \frac{dR_\pm}{d\lambda_\gamma}
    =
    \omega_\pm(H_\pm R-1).
    \label{eq:dRdlambda_photon}
\end{equation}
Therefore, if
\begin{equation}
    s_\pm = \theta R ,
    \label{eq:s_theta_R}
\end{equation}
with constant observed angular separation \(\theta\), then
\begin{equation}
    \frac{ds_-}{d\lambda_{\gamma}}\bigg{|}_- - \frac{ds_+}{d\lambda_{\gamma}}\bigg{|}_+
    =
    \theta
    \left[
        \omega_-(H_-R-1)
        -
        \omega_+(H_+R-1)
    \right]_{\Sigma}.
\end{equation}
Since \(s=\theta R\) at the wall, this agrees exactly with \eqref{eq:s_jump}. Thus the shell-induced discontinuity in \(ds/d\lambda_\gamma\) is precisely the discontinuity implied by the change in \(dR/d\lambda_\gamma\) across the wall.

We conclude that, for central radial photon bundles in this matched flat-FLRW spacetime, the angular-diameter distance is given by the areal radius,
\begin{equation}
    D_A = R .
    \label{eq:DA_equals_R}
\end{equation}
The null wall does not introduce an additional lensing correction to \(D_A\) beyond the matching of the areal radius itself. However, this statement should not be confused with the stronger claim that the function \(D_A(z)\) is identical to that of a single homogeneous FLRW cosmology. The bubble can still modify the relation between redshift, emission time, and areal radius through the different expansion histories and the matching conditions influencing the time coordinates across the wall.

\section{Angular Dependence of the Redshift to Last Scattering}

\label{append:z_theta}

\begin{figure}
    \centering
    \includegraphics[width=0.9\linewidth]{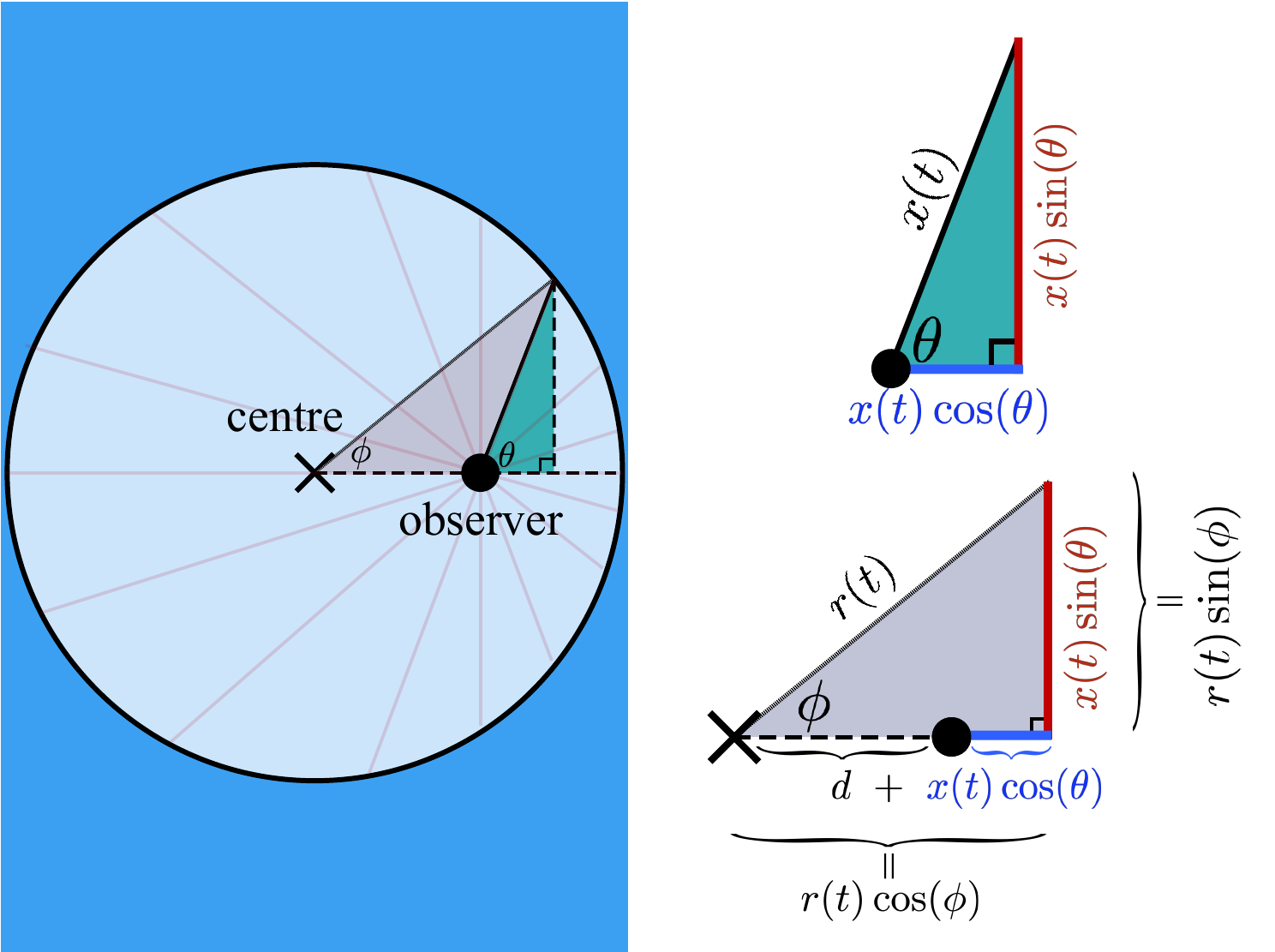}
    \caption{Illustration of trigonometric relations needed to derive the observed redshift as a function of angle $\theta$ of an off-centre observer. Assuming the CMB photons are uniformly incident upon the observer from all directions, a computation of deflection of the photons as they travel across the bubble wall, as can be seen in the derivation of $z(\theta)$ in Appendix \ref{append:z_theta}, is not needed.}
    \label{fig:z_of_theta_geometry}
\end{figure}

In this appendix we describe the procedure used to compute the redshift of CMB photons as a function of the angle at which they are observed. We consider an observer located inside the bubble at a fixed final time \(t_{-}^{0}\), corresponding to the present time in the interior cosmology. As illustrated in Fig.~\ref{fig:z_of_theta_geometry}, the observer is displaced from the centre of the bubble by a comoving distance \(d\). A photon arriving at the observer from direction \(\theta\) is followed backwards in time until it intersects the bubble wall.

We denote by \(x_-(t_-)\) the comoving distance travelled by the photon in the interior FLRW region between the observation time \(t_{-}^{0}\) and an earlier interior time \(t_-\). For a radial null geodesic in the interior geometry,
\begin{equation}
    x_-(t_-)
    =
    \int_{t_-}^{t_{-}^{0}} \frac{dt'}{a_-(t')} ,
    \label{eq:xminus_def}
\end{equation}
where \(a_-(t_-)\) is the interior scale factor. The photon intersects the wall when the endpoint of this interior geodesic lies on the bubble boundary. Let \(r(t_-)\) denote the comoving radius of the wall, written in interior coordinates. The geometry gives
\begin{align}
    r(t_-)\sin\phi &= x_-(t_-)\sin\theta ,
    \label{eq:geom_perp}
    \\
    r(t_-)\cos\phi &= d + x_-(t_-)\cos\theta ,
    \label{eq:geom_parallel}
\end{align}
where \(\phi\) is the angle subtended at the bubble centre. Eliminating \(\phi\), the wall-crossing time in the interior coordinates is determined by
\begin{equation}
    r^2(t_-)
    =
    d^2
    + x_-^2(t_-)
    + 2d\,x_-(t_-)\cos\theta .
    \label{eq:crossing_condition}
\end{equation}
For each observed direction \(\theta\), we solve this equation numerically for the interior crossing time,
\begin{equation}
    t_-^{\rm c}(\theta) .
\end{equation}
Once \(t_-^{\rm c}\) is known, the corresponding interior path length is
\begin{equation}
    x_-^{\rm c}(\theta)
    =
    x_-\!\left(t_-^{\rm c}(\theta)\right) .
\end{equation}

The exterior time at which the same photon crosses the wall is then obtained from the wall matching relation,
\begin{equation}
    t_+^{\rm c}(\theta)
    =
    t_+\!\left(t_-^{\rm c}(\theta)\right),
    \label{eq:tplus_cross}
\end{equation}
where the function \(t_+(t_-)\) is determined numerically from the wall evolution, as described in Sec.~\ref{sec:redshift}. The same numerical solution also gives the Jacobian relating the two time coordinates at the wall,
\begin{equation}
    \left.
    \frac{dt_+}{dt_-}
    \right|_{\rm c}
    =
    \left.
    \frac{dt_+}{dt_-}
    \right|_{t_- = t_-^{\rm c}(\theta)} .
    \label{eq:wall_jacobian}
\end{equation}

We take the surface of last scattering to be fixed in the exterior cosmology, at time \(t_+^{\rm ls}\). The total redshift of a photon observed at angle \(\theta\) can then be written as the product of three factors: the redshift accumulated in the exterior FLRW region, the frequency shift associated with crossing the wall, and the redshift accumulated in the interior FLRW region. This gives
\begin{equation}
    1 + z(\theta)
    =
    \frac{a_-(t_{-}^{0})}{a_-\!\left(t_-^{\rm c}(\theta)\right)}
    \left.
    \frac{dt_-}{dt_+}
    \right|_{\rm c}
    \frac{
        a_+\!\left(t_+^{\rm c}(\theta)\right)
    }{
        a_+(t_+^{\rm ls})
    } .
    \label{eq:ztheta}
\end{equation}
Here \(a_+(t_+)\) is the scale factor of the exterior cosmology. The first factor accounts for propagation from the wall to the observer inside the bubble, the final factor accounts for propagation from last scattering to the wall in the exterior spacetime, and the middle factor accounts for the conversion between the interior and exterior time coordinates at the wall.

Equation~\eqref{eq:ztheta} is evaluated independently for each observed direction. The angular dependence of the observed CMB redshift arises because photons arriving from different values of \(\theta\) cross the bubble wall at different interior times \(t_-^{\rm c}(\theta)\), and therefore at different exterior times \(t_+^{\rm c}(\theta)\). The resulting function \(z(\theta)\) is the quantity used in Sec.~\ref{sec:cmb_dipole_constraints} and Sec.~\ref{sec:ksz_constraints} to characterize the anisotropic CMB signal seen by an off-centre observer.

\section{Snell's Law for Thin-Wall Bubbles}
\label{append:snell}
\begin{figure}
    \centering
    \includegraphics[width=0.8\linewidth]{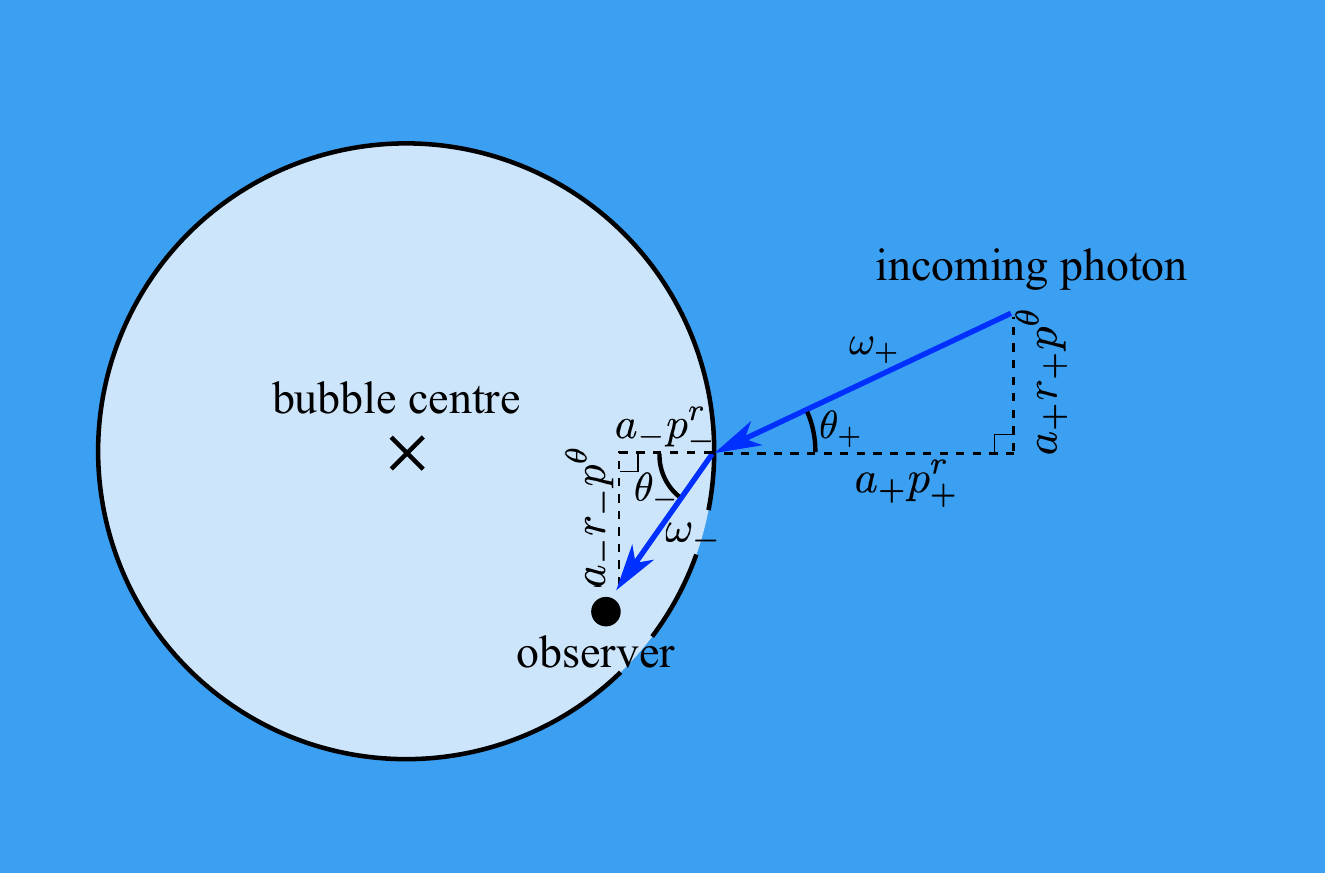}
    \caption{Illustration of photon deflection for a non-central observer as a result of the bubble cosmology. This deflection results in a bubble wall Snell's Law.}
    \label{fig:snells_law}
\end{figure}

We derive the equivalent of Snell's Law for a thin shelled bubble using the photon normalization condition.

\begin{align}
    &0 = p_\mu p^\mu = g_{\mu\nu}p^\nu p ^\mu = -\qty(p^t)^2 + a^2(t)\qty[\qty(p^r)^2+r^2\qty(p^\theta)^2]\\
    &\implies \abs{p^t} = a(t)\sqrt{\qty(p^r)^2+\qty(rp^\theta)^2} = \omega.
\end{align}
Therefore we see that
\begin{equation}
    \sin\theta = \frac{arp^\theta}{\omega}.
    \
\end{equation}
$p^\theta$ is the angular component of the momentum, and as illustrated in Fig. \ref{fig:snells_law}, is parallel to the bubble wall and therefore is not impacted by the photon crossing into the bubble. The radial component of the momentum is impacted and therefore leads to the change in the frequency, as explored in the paper as the change in redshift, and a change in the angle of the photon with respect to the wall.

Note that the continuity of physical distance, $a_\pm r_\pm$, across the bubble boundary leads to the equality $a_-r_-p^\theta = a_+r_+p^\theta$ at the boundary. Therefore, we obtain
\begin{equation}
    \omega_+ \sin\theta_+ = a_+r_+p^\theta = a_-r_-p^\theta = \omega_-\sin\theta_-,
\end{equation}
giving us the exact form of Snell's Law:
\begin{equation}
    \omega_+\sin\theta_+ = \omega_-\sin\theta_-.
\end{equation}
From the redshift jump at the bubble wall, we know that $\omega_+ > \omega_- \implies \frac{\omega_+}{\omega_-}>1$. Therefore we see that $\frac{\sin\theta_-}{\sin\theta_+
}>1$. Given the problem setup where $0\leq \theta_-,\theta_+<\pi/2$, we can determine that $\theta_->\theta_+$. Therefore, null rays crossing into the bubble will always bend away from the normal, magnifying distant objects to the inside observer. This is the same effect as light traveling from a medium with a higher refractive index in which it travels more slowly to a medium with a lower refractive index in which it travels more quickly. Light traverses comoving space more slowly in the cosmology with more dark energy, and traverses comoving space more quickly in the cosmology with less dark energy.

\begin{figure}
    \centering
    \includegraphics[width=0.9\linewidth]{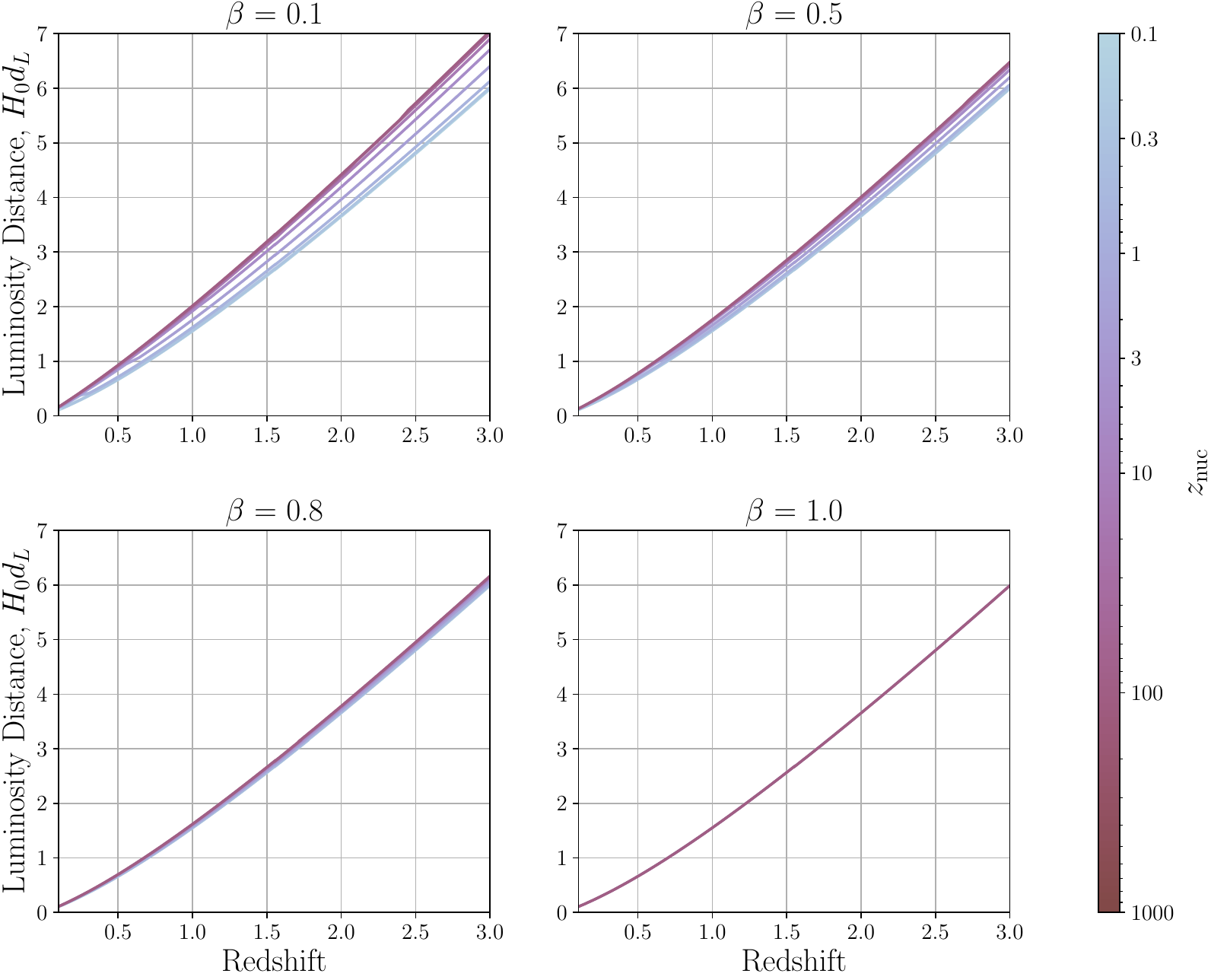}
    \caption{Luminosity distance-redshift relationship in a bubble cosmology. The relationship for varying $\beta$ and $z_{\rm nuc}$ values is plotted.}
    \label{fig:luminosity_dist}
\end{figure}

\section{Luminosity Distance and Distance Modulus}
\label{append:luminosity_distance}

The presence of the bubble changes the relation between emission time, observed redshift, and distance, but it does not change the operational definition of the luminosity distance. For a source with bolometric luminosity $L$ and observed bolometric flux $F$, the physical luminosity distance $D_L$ is defined by
\begin{equation}
    F \equiv \frac{L}{4\pi D_L^2}\, .
\end{equation}

If photon number is conserved along the ray bundle connecting the source and the observer, the luminosity distance is related to the angular-diameter distance by Etherington's distance-duality relation,
\begin{equation}
    D_L = (1+z_{\rm obs})^2 D_A .
\end{equation}

This relation is not a consequence of the FLRW metric. It holds in general metric spacetimes for photons propagating along null trajectories, provided that photons are neither absorbed, emitted, nor scattered out of the beam. In the present model, the bubble matching can change the photon trajectory, the observed redshift, and the angular-diameter distance, so both $z_{\rm obs}$ and $D_A$ must be computed along the actual path through the bubble spacetime.

For the radial rays considered here, we denote the dimensionless angular-diameter distance along the ray by $d_{\rm phys}$. The dimensionless luminosity distance is therefore
\begin{equation}
    d_L =
    d_{\rm phys}\left(1+z_{\rm obs}\right)^2 .
    \label{eq:dL_bubble}
\end{equation}
This expression has the same form as in a homogeneous cosmology, but the quantities appearing in it are those computed in the bubble spacetime. In particular, both $d_{\rm phys}$ and $z_{\rm obs}$ generally differ from their values in a single homogeneous FLRW universe. The luminosity distance-redshift relation is plotted in Fig.~\ref{fig:luminosity_dist}.

The physical distance modulus is
\begin{equation}
    \mu
    =
    5\log_{10}\left(\frac{D_L}{10\,{\rm pc}}\right).
\end{equation}
In terms of the dimensionless luminosity distance $d_L=(H_0/c)D_L$, this can be written as
\begin{equation}
    \mu
    =
    5\log_{10} d_L
    +
    \mu_0 ,
    \label{eq:distance_modulus_dimensionless}
\end{equation}
where
\begin{equation}
    \mu_0
    =
    5\log_{10}\left(\frac{c/H_0}{10\,{\rm pc}}\right)
\end{equation}
is a constant offset. For uncalibrated supernova samples, this offset is degenerate with the absolute magnitude calibration and with $H_0$.

Thus, when comparing to a fiducial cosmology, we use
\begin{equation}
    \Delta \mu
    \equiv
    \mu-\mu^{\rm fid}
    =
    5\log_{10}
    \left(
        \frac{d_L}{d_L^{\rm fid}}
    \right)
    + {\rm const.}.
    \label{eq:delta_mu}
\end{equation}
A different value of $H_0$ corresponds only to a vertical shift of the Hubble diagram, while the redshift dependence of $\Delta\mu$ captures the physical effect of the bubble. As noted in Ref. \cite{DESI_DR2} for the supernova samples of Refs. \cite{Pantheon+, DESY5, Union3}, this additive vertical offset is unconstrained.

\section{Qualitative Comparison with Supernovae}
\label{append:sne_comparison}

\begin{figure}[t]
    \centering
    \includegraphics[width=0.9\linewidth]{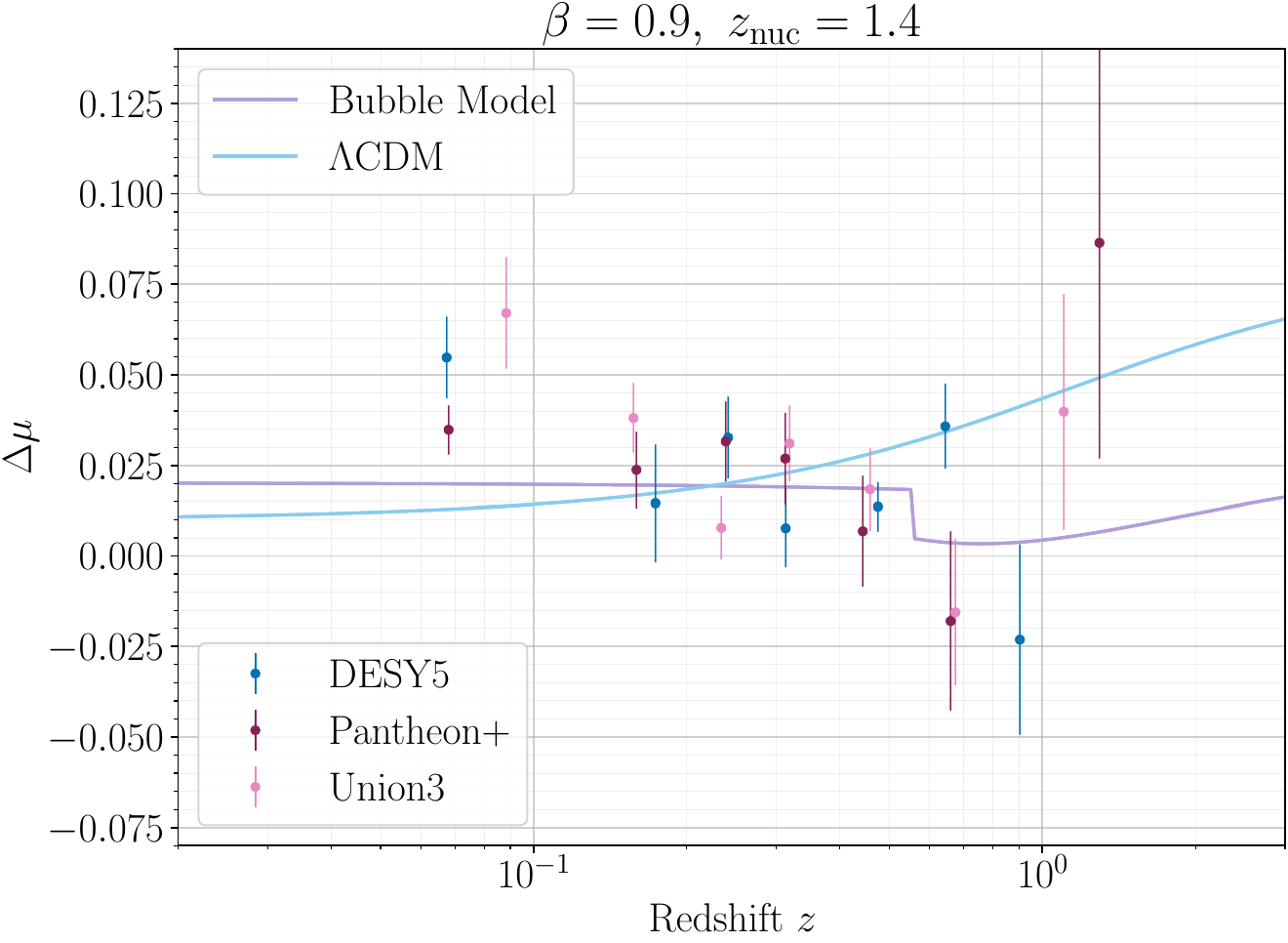}
    \caption{
    Illustrative comparison of distance-modulus residuals for the bubble model with $\beta=0.9$ and $z_{\rm nuc}=1.4$ against the Union3, Pantheon+, and DESY5 Type~Ia supernova compilations. The purple curve shows the bubble prediction, while the blue curve shows the corresponding no-bubble reference model. Constant vertical offsets have been applied to the model curves and to the supernova residuals, since the absolute normalization of the supernova Hubble diagram is degenerate with the supernova absolute magnitude and with $H_0$. The comparison is intended only to show the approximate redshift dependence of the residuals, $\Delta\mu$. No supernova likelihood is used, and no constraints are inferred from this figure.
    }
    \label{fig:sne}
\end{figure}

We also show a qualitative comparison between the luminosity-distance prediction of the bubble model and recent Type~Ia supernova compilations; see Fig.~\ref{fig:sne}. This comparison is not used to fit the model parameters and is not included in the constraints quoted in the main text. The purpose of this appendix is only to illustrate the approximate redshift dependence of the bubble-induced distance-modulus residuals, for readers interested in how the model would appear in a supernova Hubble diagram, as the vertical offset is unconstrained.

A fully self-consistent supernova analysis in the bubble spacetime would require more than evaluating the luminosity distance to the supernovae themselves. The observed supernova distance moduli depend on light-curve standardization, calibration of the absolute magnitude, selection effects, and correlated statistical and systematic uncertainties. In addition, in the present model, the nearby anchors used in the distance ladder, the calibrator supernovae, and the Hubble-flow supernovae need not all occupy the same region of the bubble spacetime. For example, sufficiently local objects could lie inside the bubble, while some of the supernovae used to determine the Hubble diagram may lie outside it or have photon trajectories that cross the bubble boundary. The usual absolute-magnitude calibration would therefore have to be recomputed within the bubble model before the supernova samples could be used as a quantitative likelihood.

This comparison is also not central to the main conclusion of this work. As discussed in the main text, the strongest restriction on the bubble model comes from the CMB. The region of parameter space allowed at the $1\sigma$ level corresponds to bubbles whose effects are too small to be visible in the DESI distance measurements, and hence the model does not provide a viable explanation of the DESI--CMB tension. The supernova comparison shown here is included as a first indication of the qualitative behaviour that could be explored in a future analysis with the full supernova likelihoods and a self-consistent treatment of the distance ladder.

\end{document}